\documentclass[11pt,a4paper]{article}

\usepackage{natbib}
\usepackage{url}
\usepackage{booktabs}
\usepackage{siunitx}
\usepackage{float}
\usepackage{graphicx}
\usepackage{amsmath,amssymb}
\usepackage{hyperref}

\title{Evaluating the Quality of Open Building Datasets for Mapping Urban Inequality: A Comparative Analysis Across 5 Cities}

\author{
  Franz Okyere$^{1,*}$\thanks{Corresponding author: franz.okyere.doktoranden@thws.de} \and
  Meng Lu$^{2}$ \and
  Ansgar Brunn$^{1}$
}

\date{} % leave empty or put a submission date

\newcommand{\affil}[2]{\noindent$^{#1}$~#2\\}
  
\begin{document}
\maketitle

\affil{1}{Faculty of Plastics Engineering and Surveying, Technical University of Applied Sciences Würzburg-Schweinfurt (THWS), 97070 Würzburg, Germany; Ansgar.Brunn@thws.de}
\affil{2}{Faculty of Biology, Chemistry \& Earth Sciences, University of Bayreuth, Germany; Meng.Lu@uni-bayreuth.de}

\vspace{1em}
\noindent \textbf{These authors contributed equally to this work.}

\newcommand{\keywords}[1]{%
  \vspace{2mm}\noindent\textbf{Keywords: } #1
}

% The commands \thirdnote{} till \eighthnote{} are available for further notes

%\simplesumm{} % Simple summary

%\conference{} % An extended version of a conference paper

% Abstract (Do not insert blank lines, i.e. \\) 

\begin{abstract}While informal settlements lack focused development and are highly dynamic, the quality of spatial data for these places may be uncertain. This study evaluates the quality and biases of AI-generated Open Building Datasets (OBDs) generated by Google and Microsoft against OpenStreetMap (OSM) data, across diverse global cities including Accra, Nairobi, Caracas, Berlin, and Houston. The Intersection over Union (IoU), overlap analysis and a positional accuracy algorithm are used to analyse the similarity and alignment of the datasets. The paper also analyses the size distribution of the building polygon area, and completeness using predefined but regular spatial units. The results indicate significant variance in data quality, with Houston and Berlin demonstrating high alignment and completeness, reflecting their structured urban environments.  There are gaps in the datasets analysed, and cities like Accra and Caracas may be under-represented. This could highlight difficulties in capturing complex or informal regions. The study also notes different building size distributions, which may be indicative of the global socio-economic divide.  These findings may emphasise the need to consider the quality of global building datasets to avoid misrepresentation, which is an important element of planning and resource distribution.
\end{abstract}

% Keywords
\keywords{open buildings dataset, AI-generated datasets, quality assessment, completeness, relative positional accuracy, deprived regions }

\begin{figure}[H]
\centering
\includegraphics[width=12cm]{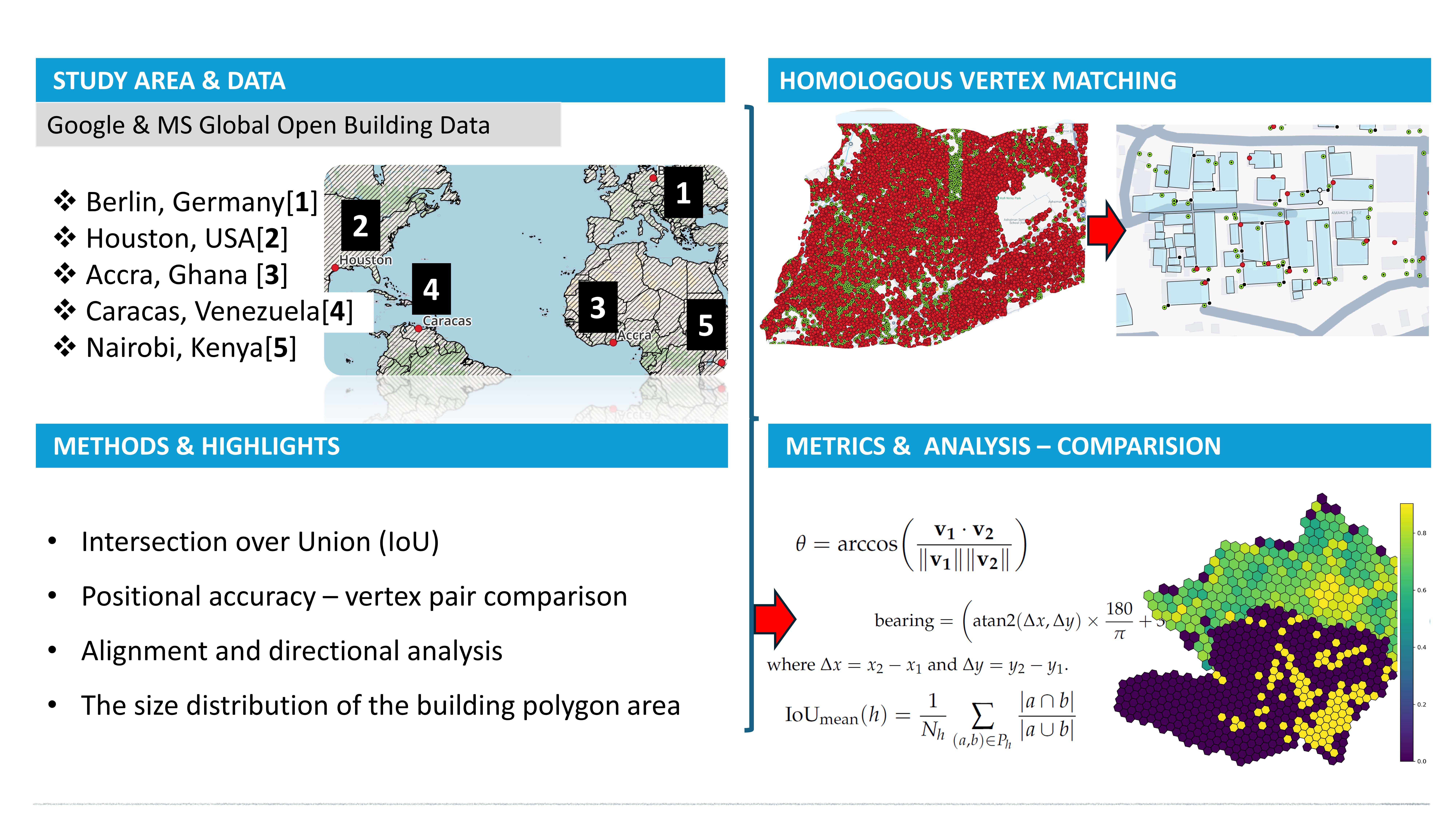}

\label{figx}
\end{figure}

\section{Introduction}

 "The informal settlements are the most accelerated form of urban growth on the planet, but they are simultaneously zones of exception—off the map, off the grid, and off the record." \citep{Davis2006}  Urban areas comprise formal and informal areas and the formal areas may be planned, as expected according to some accepted scheme, standard or policy. While the informal settlements, also known as slums, favelas, and ghettos, are deprived regions where the poor live. The contrast between the two lies in the apparent lack of essential amenities and associated health issues due to the dirty environments \citep{Thomson2020,Zeppelini2021,Patriciaa2023,Kuffer2023}. In Ghana, 5.5 million people lived in slums in 2017,and 8.8 million in 2020 \citep{GhanainTimes2024}. Since these areas suffer from inequality in security of tenure, the lack of attention may lead to neglect in spatial data acquisition by government agencies.

To support urban applications such as slum dynamics and urban planning, accurate data is required. This has been obtained from traditional methods such as photogrammetry, field surveys and remote sensing \citep{Trotter1991,Longley2001,Mahabir2018}. Accurate, timely and frequent data are essential to map deprived regions, as indicated by \cite{Engstrom2019, Gruebner2014, Kuffer2016}. The usage of traditional data, especially remote sensing data, can support a global study of slums. Nonetheless, the use of remote sensing data can only be economical processing further. Due to the high cost and time-consuming nature of data acquisition \citep{Chelanga2022,Thomson2020}, satellite imagery need to be cost-effective  \cite{Kohli2021} if used. The efficient use of satellite imagery by application of artificial intelligence, together with remotely sensed images, results in the generation of new data. AI encompasses two broad methods or concepts: machine learning (ML) \citep{Sheykhmousa2020, Adugna2022} and deep learning \citep{Bisong2019}. Using deep learning \citep{Bisong2019, Ray2019, Janiesch2021} to generate spatial data accurately, almost automatically, can reduce the tedium of spatial data acquisition and serve as a cost-effective alternative to conventional data acquisition methods \citep{Schwartz2021, Chamberlain2024,Foody2024}. Artificial intelligence (AI) models are built using bespoke algorithms that train, validate, and test labelled data. Once built, the model makes inferences \citep{Mohanty2020} from unseen data during the testing phase. With the advent of open building footprint datasets, the accuracy of these automatically generated datasets needs to be assessed. Notwithstanding, the models are evaluated for precision as part of training, validation and testing phases \citep{Farhadpour2024, Maxwell2021}. The attention received by AI-generated datasets requires that quality issues should be explored. There are many models and approaches and data generated are being made available online \citep{Sirko2021}. AI, ML-generated data can be integrated into official datasets when proper validation of data are not implemented as part of standard practices\citep{Hansen2023}. 

Opportunities in using AI for national mapping were investigated by \citep{Murray2020}. This is as a result of the leverage provided by open data, which is accessible, editable, and shareable, promoting government transparency, innovation, and collaboration \citep{reggi2016,Morelli2019}. Open data does not mean we must condone inaccurate data, especially for application areas where precision and accuracy are paramount.  GI derived from open datasets has the potential to support decision-making; however, automatic techniques should be used with caution due to potential inherent imprecision. Reviewed and validated data may be integrated into mainstream Volunteered Geographic Information (VGI) \citep{Roth2020, Enami2023}. VGIs serve as an example of open data and are compiled from community and individual contributions, supported by the theory of citizen science \citep{Haklay2015,FonteCidaliaCosta2017}, and made freely available. But how quality is defined in the context of these open datasets remains interesting. 

\cite{Bahl2015} defines data quality by the following categories: accuracy, completeness, consistency, credibility, and currentness. Most research focuses on one aspect of quality- for example quality assessment of VGI based on completeness \citep{Wang2020, Salvucci2021, Chen2023, Ullah2023}. Studies on quality assessment compare the OSM with reference datasets based on semantics and spatial attributes. Intrinsic and extrinsic quality assessment is helpful in the determination of the extent to which a dataset represents the real-world \citep{Barron2014, Zhang2017, Jacobs2020, Yamashita2023}.   While intrinsic evaluation looks at data and its metadata, extrinsic assessment depends on reliable ground truth data from national surveys \citep{Jacobs2020,Madubedube2021}.  Most ground truth data may not be readily available, making the pursuit of alternative data sources a worthy cause; the assessment is then subject to the accuracy of the assumed reference dataset.  While satellite imagery faces challenges like resolution and cloud cover, it still has much to offer regarding geographic information analyses \citep{Liu2019c}.  Even though OBDs are mostly generated from  VHR imagery using deep learning, creating new vector datasets \citep{Touzani2021, Hoin2022}, few studies assess the quality of these datasets, especially Google and Microsoft global building datasets like the \cite{Gonzales2023} and \cite{ Gevaert2024}. Research on the quality and the rules for modelling building polygons is scanty, though methods for comparing VGI data exist \citep{Xie2015, Zhou2022, Liu2023a}.

Automatically extracted buildings can be used for the analysis of neglected, deprived regions. While the benefits of using OBDs are obvious, it is interesting to look at the variances in data quality, taking an intra-urban approach that considers the country-side too. This will help make comparisons that can reveal assurance of an equitable modelling and representation of geographic features, irrespective of the side of the globe. \cite{Soman2020} used methods that allowed them to pinpoint infrastructure deﬁcits, detecting informal settlements.

This research evaluates the quality of open building datasets, based on those used in \cite{Zhou2022} and \cite{Biljecki2023} and attempt to answer the following research questions:

\begin{enumerate}
     \item What methods can be employed to assess the quality of building datasets derived from very high-resolution imagery using AI, and how do these assessments ensure equitable geographic feature representation across diverse regions?
     
     \item To what extent can the building footprints detected by Google and Microsoft AI models for building extraction be comparable?

     \item How can a measure of similarity in the Open Building Dataset be measured and visualised across cities in the global South and North?

 \end{enumerate}

\subsection{Study Regions} 
\subsubsection{Study Site Selection}
The purposeful choice of research sites is meant to showcase urban inequality based on geographical and economic context. Selected as representative cities from the Global North, Houston and Berlin are chosen for their wealth and organised urban settings, which fit the formal status; morphological slums were indexed this way by \cite{Taubenbock2018}. On the Global South, Accra, Nairobi, and Caracas are chosen based on their lower country GDPs and the predominance of informal settlements. This selection guarantees a fair representation of several urban morphologies. This enables an assessment of the quality and prejudices of OBDs in capturing urban morphology in deprived and developed cities. The spatial distribution of these study areas allows for a comparison of the datasets, hence improving generalisability. 

The study regions are made up of three cities—Accra, Nairobi, and Caracas. Houston in North America and Berlin in Germany are included for the above-stated reasons. Cities in the global South are characterised by informality, while cities in the global North are mostly associated with affluence and have formal, planned urban built-up areas. Understanding the economic context of the study region is important, and the Gross Domestic Product (GDP), which serves as an indicator of a country's wealth, can support the differentiation between the selected study areas \citep{WBG2025}. The global South and global North ideas are adapted; usually situated to the north of less developed countries, the countries of the world marked by high levels of economic and industrial growth are referred to as the global North countries and vice versa.  The Greater Accra region, which hosts the capital of Ghana, Accra, has a population of approximately 5.4 million as of 2021 \citep{Borsah2025}, and Ghana's GDP in 2023 was around \$76 billion. 
Caracas, Venezuela's capital, has a metropolitan population of about 2.9 million in 2021, and Venezuela's GDP is approximately \$482 billion in 2023. Nairobi, the capital of Kenya, has a population of about 4.7 million and a country GDP of around \$108 billion in 2023. Berlin, Germany's capital, has a country GDP of approximately \$4.5 trillion. Harris County, Texas, which includes Houston, has a population of about 4.8 million, according to the 2020 Census, with the USA's GDP in 2023 estimated at \$27 trillion. 

\section{Related Works}

Open datasets are increasingly being considered for spatial data management and a diverse range of application areas \citep{Liu2014, Roth2020}. For example, \citep{Sirko2021} created an open structures dataset with 516 million footprints across Africa through a model training pipeline. These datasets support applications such as energy \citep{Berezina2022}, hurricane damage \citep{Enami2023}, socioeconomic \citep{Lim2021}, and land use \citep{Trigaux2017, Nematchoua2020, Nematchoua2020a, Susca2020} research. The advent of open datasets is justified by the high cost of data, which is a very important aspect of GISs \citep{Longley2001} and the availability of various methods in computer science and related fields. AI for generating open data constitutes merely a segment of the overall data generation pipeline. Post-processing techniques are integrated into the techniques used.

Methods used in post-processing the feature extraction abound. They range from distance regression with boundary classification that produces well-defined contour maps. This can be improved further to retain image feature information \cite{Xu2024}. Others look at Multi-task Edge Detection (MTED) which enhances vectorised building outline accuracy \cite{Wu2023} by means of geometry-guided post-processing and rotational bounding box detectors. Models such as the multitask Res-U-Net with attention mechanisms faithfully extract building roofs and forms \cite{Li2022}. Other approaches include the use of Digital Elevation Models (DEMs) from LIDAR data and self-supervised learning(SSL) that improves the distinction between bare Earth and buildings \cite{Vats2024}. Furthermore, \cite{Li2024} refined high precision building footprints using bayesian algorithms. Strategies for data augmentation enhance dataset correctness and resilience \cite{Li2022}. Using partial data from OpenStreet Map (OSM), semi-supervised learning improves model variety and robustness for large-scale extraction \cite{Huang2024}. How good is the AI-detected open datasets, bearing in mind that open nature could also connote a compromise on precision and accuracy? 

Assessment of the quality of OBDs require much research to gain insight into potential usage of the result of the feature extraction  \cite{Ren2017, Xia2023} process and post-processing. Spatial data quality goes beyond the positional accuracy \citep{John1993} and touches on thematic and temporal accuracy, completeness, logical consistency and usability\citep{FonteCidaliaCosta2017, Stehman2018}. However, open dataset quality research doesn't always cover all aspects, focusing on a few \citep{Medeiros2019}. Completeness should be categorised into attribute completeness, as in \cite{Biljecki2023}, and spatial completeness, which our study considers; the presence of buildings is checked against a reference layer.

For example, \citep{Brovelli2018} used an extrinsic approach to evaluate OSM building footprint accuracy and completeness, comparing data with reference datasets in Lombardy, Italy. Similarly,  \citep{Goldberg2018, Zhuetal2021, Hoin2022} highlighted precise footprint identification using orthorectified data and digital surface models from satellites. \citep{Zhou2022} also evaluated OBD buildings' global quality, revealing a strong correlation with reference completeness. Also, novel spatial sample approaches have been used to evaluate remote sensing-derived products \citep{Xie2015, Liu2023a}. While there's increasing interest in evaluating open building footprint dataset quality, the focus has been on spatial or positional accuracy and completeness. 

%insert something on bias

%bias
While morphological slums analyses \citep{Taubenbock2018, Jochem2021} require building footprint data often display biases towards formal areas \citep{Gevaert2024}. Omissions and clustering have been identified in datasets from Google and Microsoft \citep{Kuffer2023}. Although global or multi-country studies evaluate Google's and Microsoft's building datasets \citep{Minghini2017, Herfort2023}, their district-level impact is less explored. Google (\href{https://sites.research.google/open-buildings}{Google open buildings}), Microsoft (\href{https://www.microsoft.com/en-us/maps/building-footprints}{Microsoft building footprints}), and Ecopia (\href{https://www.ecopiatech.com/global-feature-extraction}{global feature-extraction}) datasets have been evaluated by \citep{Chamberlain2024} . \cite{Chamberlain2024, Gevaert2024} compared building counts and a similarity index across countries using global datasets. The focus of such studies are not on the potential inequities or bias of the dataset as a result of the performance of underlying models in diverse geographic contexts.

%insert something on bias
% What is bias and inequality in the AI-gen dataset? How does the quality relate?
AI bias affect the resultant data product and come from biased data and algorithms, which can generate systematic inequality, therefore influencing the quality and reliability of systems \citep{IMB2023, Schwartz2022, Nazer2023}. Biases in datasets or algorithm design can spread erroneous results, hence supporting discrimination \citep{Hanna2025}. While compromising audit quality and risk identification \citep{Ferrara2024, murikah2024}. This might severely affect areas including recruiting, lending, criminal justice and many application areas.  However, fair and reliable artificial intelligence depends on the ability to address bias.  Due to misallocation of resources in underdeveloped areas, biased datasets can under-represent vulnerable populations in the generation of geographic data  \citep{Gevaert2024, Wang2019a}.  Research suffers from often lacking dataset uncertainties, including the visibility of rooftops and sample size.  For example, \cite{Nurkarim2023} enhanced Paris building detection by 74.66\% 
 against 56.19\% in Khartoum.

%%%%%%%%%%%%%%%%%%%%%%%%%%%%%%%%%%%%%%%%%%
\section{Materials and Methods}

\subsection{Data Sources} 
 Google and Microsoft datasets are used in our evaluations, and positional accuracy and completeness metrics used are similar to those used by \cite{Brovelli2018}. Although Google’s data enhances availability in low- and middle-income countries \citep{Gonzales2023}, its accuracy at the district level needs further investigation in both the global South and global North.  Google open building datasets are available on their website, as a collection,  \href{https://developers.google.com/earth-engine/datasets/catalog/GOOGLE_Research_open-buildings_v3_polygons}{here} and can be custom downloaded using a JavaScript Application Programming Interface(API) \citep{GoogleEE2024}. Microsoft's datasets are openly available \href{https://github.com/microsoft/GlobalMLBuildingFootprints}{here}, but data can be downloaded for the global South countries and the narrow fringes of their borders with other continents. Building footprint data for Berlin was obtained from the \href{https://github.com/microsoft/GlobalMLBuildingFootprints}{VIDA} group at Microsoft Research. This dataset provides harmonised building outlines based on both Google and Microsoft Open Buildings data. However, since Microsoft does not provide building data for Berlin through its JavaScript API or official GitHub repository, the VIDA platform served as the primary source for this study area. The data is accessed through their interactive web interface, which enables spatial queries and downloads for selected regions \citep{MSGoogleVIDA2024}.

\begin{figure}[H]
\centering
\includegraphics[width=12cm]{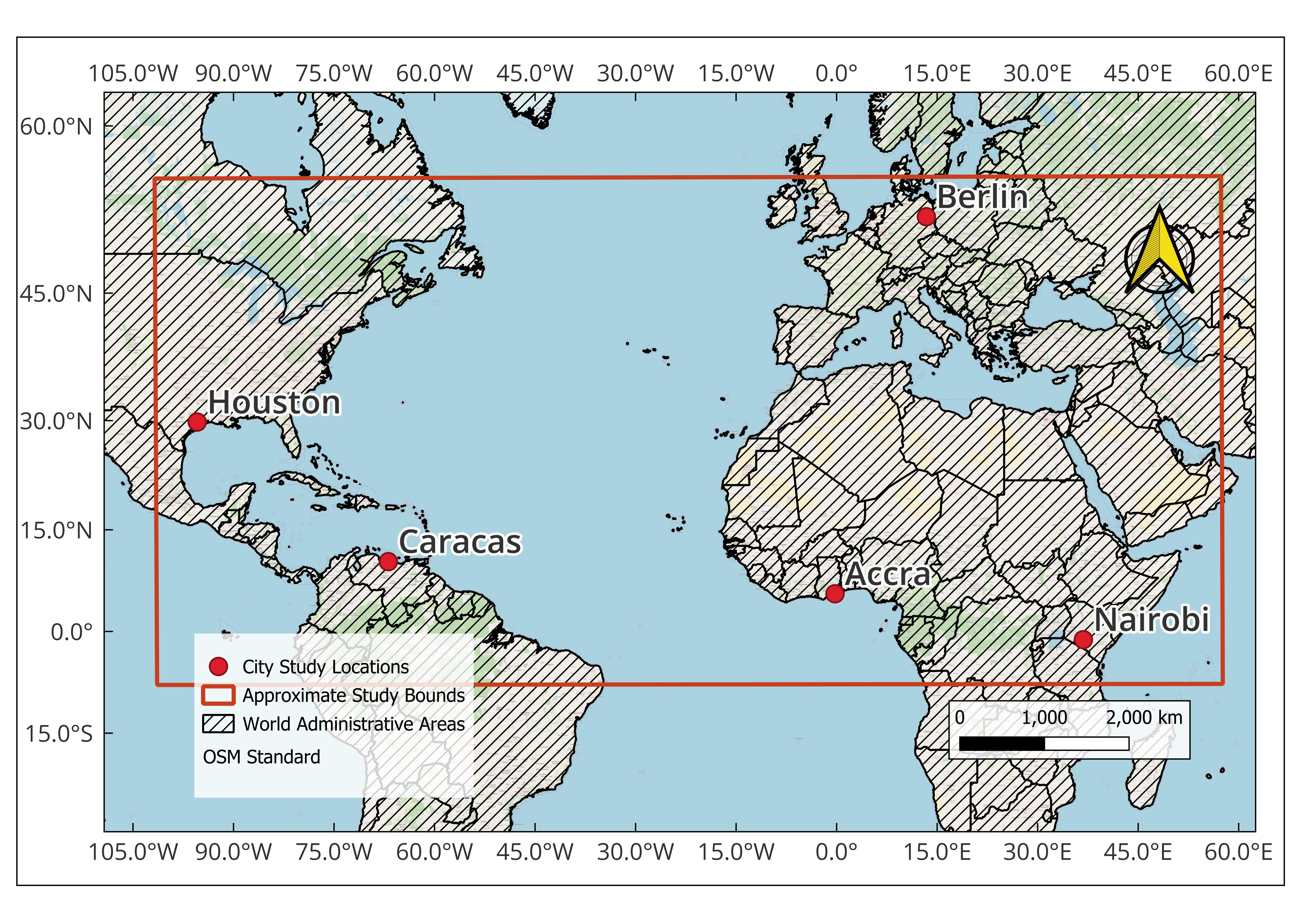}
\caption{Map of the study region showing the city locations of the study. From east to west: Houston, Caracas, Accra, Berlin and Nairobi.} 
\label{fig2}
\end{figure}

Metropolitan administrative boundaries are used to define each study region. Even though not every area in the study boundary is urban, the study can ignore the effect at the global scale. Countrysides are usually characterised by low building density from a global view.  For Houston, this study uses the Harris County boundary, adapts the Greater Accra region boundary for Accra, uses the Berlin state boundary to define the study area of Berlin, and finally, uses the Caracas and Nairobi district polygons for our analysis.

%# # # # # # # # # # # # # # # 

%insert the code and methodology here
%\subsection{ Algorithm} 
\subsection{Positional Accuracy based on homologous points—processing and transformation}

This subsection outlines the methodology for processing and transforming geospatial data using affine transformations. The process involves computing angles and bearings to identify the vertices matching two datasets. The goal is to get at least five well-distributed points for an affine transformation. Four of these points are chosen to be located around the four corners of the dataset's minimum bounding envelope., The fifth point is placed at the centre of the convex polygon, which contains all the building vertices; this is done for all datasets for all study cities.

\subsubsection{Affine Transformation Parameters}

 At least three matching points from two datasets are required to compute the affine transformation parameters. The vertices are denoted as $\{(x_i, y_i)\}$ and $\{(x'_i, y'_i)\}$ for the source and destination points, respectively. The relationship is expressed in matrix form as follows:

\begin{equation}\label{eq:affine-transformation}
\begin{bmatrix}
x'_i \\
y'_i \\
\end{bmatrix}
=
\begin{bmatrix}
a & b & t_x \\
c & d & t_y \\
\end{bmatrix}
\begin{bmatrix}
x_i \\
y_i \\
1 \\
\end{bmatrix}
\end{equation}

Parameters $a, b, c, d, t_x, t_y$(Equation \ref{eq:affine-transformation}) can be computed using least-squares fitting over the corresponding points in the compared datasets;  when more than the minimum number of points are used, the least-squares solution can be applied. Applying least squares ensures that coordinates are more harmonised and a fair comparison between the coordinates. This approach from the previous research is adapted, using a minimum of five well-distributed vertex points, despite the fact that an affine transformation requires a minimum of three coordinates. The choice of this number should be carefully considered, as it has the potential to impact the computations of relative positional accuracy. The more points are added, the more rigorous the solution, but overfitting can occur.

\subsubsection{Residuals and Goodness of Fit}

The goodness of fit is assessed using the residuals from the least-squares solution. The residual vector $\mathbf{r}$ is calculated, and the fit error $E_\text{fit}$ is as shown in equation \ref{eq:fiterror}.

\begin{equation}\label{eq:fiterror}
E_\text{fit} = \sqrt{\sum_{i} r_i^2}
\end{equation}
where $r_i$ represents the $i^{th}$ residual for $i^{th}$ vertex.
The fit error is not reported here because it also relates to relative accuracy. The room can be allowed for the research question to be answered by making the following assumption. The work makes the assumption that scientific inquiry is predicated on the notion of variable control, whereby one factor is manipulated. In contrast, all other variables remain constant to determine causal relationships and ensure meaningful comparisons \citep{McMillan2010}.

\subsubsection{Angle and Bearing Calculations}

Equation \ref{eq:anglecomputation} is used to compute the angle between three consecutive points $P_1$, $P_2$, and $P_3$, with $P_2$ at the centre, the angle $\theta$. 

\begin{equation}\label{eq:anglecomputation}
\theta = \arccos\left(\frac{\mathbf{v_1} \cdot \mathbf{v_2}}{\|\mathbf{v_1}\| \|\mathbf{v_2}\|}\right)
\end{equation}

Where $\mathbf{v_1}$ and $\mathbf{v_2}$ are vectors defined by:

\[
\mathbf{v_1} = \begin{bmatrix} x_1-x_2 \\ y_1-y_2 \end{bmatrix}, \quad \mathbf{v_2} = \begin{bmatrix} x_3-x_2 \\ y_3-y_2 \end{bmatrix}
\]

The bearing between two points $P_1$ and $P_2$ is computed using equation \ref{eq:bearing} as seen below.

\begin{equation}\label{eq:bearing}
\text{bearing} = \left(\operatorname{atan2}\left(\Delta x, \Delta y\right) \times \frac{180}{\pi} + 360 \right) \mod 360
\end{equation}

where $\Delta x = x_2 - x_1$ and $\Delta y = y_2 - y_1$.

\subsubsection{ Mean Error Calculation}

The mean of angle differences, bearing differences, and positional shift is calculated as:

\[
\mathrm{Mean} = {\frac{1}{n} \sum_{i=1}^{n} (\Delta_i)}
\]

Where $\Delta_i$ represents the difference in angles, bearings, or positions between the matched vertices.

\subsection{Completeness Analysis}

This subsection describes the methodology used to analyse the completeness of spatial data by comparing overlapping datasets of buildings. The initial results indicate a wide margin between building counts of OSM buildings and the OBDs (see the overlap data in Tables \ref{tab:microsoft_overlap_metrics} and \ref{tab:google_overlap_metrics} and the distribution of overlapping and non-overlapping buildings in the appendix \ref{sec:overlap-analysis}).

\subsubsection{Data Preparation }

The analysis utilises the Microsoft Building Footprints and OBD building layers as input datasets. For fairness, though computationally expensive, all features in the dataset are evaluated to ensure they are valid polygons. Since a comparison of datasets should be with respect to a common reference frame, the data needs to be projected into a common Coordinate Reference System (CRS). A CRS suitable for area calculations is used, specifically using the Universal Transverse Mercator (UTM) \cite{Karney2011, Kennedy2000} projection for all cities to ensure consistency. The UTm zones used are as follows: Accra, UTM Zone 31N; Caracas, UTM Zone 19N; Nairobi, UTM Zone 37S; Berlin, UTM Zone 33N; and Houston, UTM Zone 15N.

All datasets, including a grid of hexagonal cells, are reprojected to this target CRS to ensure accurate spatial joins and area calculations.

\subsubsection{Spatial Joining and Area Calculations}

To examine metrics for quality assessment, buildings from both datasets are spatially joined to a grid of contiguous hexagonal cells using a spatial intersection:

\begin{itemize}
    \item Let \( A \) denote the OSM building layer.
    \item Let \( B \) denote an OBD layer.
    \item Let \( H \) denote the hexagonal grid.
\end{itemize}

The spatial join operation allows buildings to be related to the hexagons they intersect, enabling the calculation of the total building area within each hexagon.

Let \( A \) denote the OSM building layer, \( B \) denote the Microsoft OBD building layer, and \( H \) denote the set of hexagonal cells. For each hexagonal cell \( h \in H \), define the mean IoU as:

\begin{equation}
\text{IoU}_{\text{mean}}(h) = 
\frac{1}{N_h} \sum_{(a, b) \in P_h} 
\frac{|a \cap b|}{|a \cup b|}
\end{equation}

where:
\begin{itemize}
  \item \( P_h = \{(a, b) \mid a \in A, b \in B, a \cap b \cap h \neq \emptyset \} \) is the set of overlapping polygon pairs from layers \( A \) and \( B \) that intersect cell \( h \),
  \item \( |a \cap b| \) denotes the area of intersection between polygons \( a \) and \( b \),
  \item \( |a \cup b| \) denotes the area of their union,
  \item \( N_h = |P_h| \) is the number of such intersecting pairs in cell \( h \).
\end{itemize}

The resulting value \( \text{IoU}_{\text{mean}}(h) \) is assigned as an attribute to the intersecting hexagon\( h \).

\subsubsection{Completeness and Jaccard Score}
 The method used by \cite{Hecht2013}, computes completeness, C, as the ratio of the area of OBD to REF buildings within a defined spatial unit. This measures how complete the OBD data is. Our work uses hexagon grids based on an initial visual inspection, choosing 2km (Accra and Houston) or 500m (Caracas, Nairobi and Berlin) apothem hexagons to fit the analysis needs. While covering whole areas fully \citep{Hecht2013}, hexagons provide a near match to circles. But given the situation where there can be more OBD buildings than OBD data, \citep{Tornros2015} cautions that the technique might exaggerate completeness. \cite{Brovelli2018} then suggests three rates: False Positive (FP), True Positive (TP), and False Negative (FN) rates. TP represents overlap between OBD and REF; FP marks missing REF areas; FN represents missing OBD areas. In this work, FN, FP and TP are not computed directly, but the Jaccard score is computed for matched polygons, which implicitly uses these values. Completeness is then assessed by how well each hexagon is covered by the OBD area, using OSM as a reference. There is a direct link between the completeness measure used in \cite{Hecht2013} and the Jaccard score. 

Particularly in tasks like feature extraction \citep{Forstner1993, Guyon2006, Ren2017, Xia2023}, image segmentation \citep{sirko2023}, the Jaccard score, as mentioned before, also known as the Intersection over Union, is used. It is widely employed to assess the similarity between two sets \citep{aryal2023,kong2022}.  It is stated as the intersection area between two polygons divided by the spatial union of the two.  Within a classification framework, it can be stated in terms of False Negatives (FN), False Positives (FP), and True Positives (TP). Unique identifiers for each polygon ensure that overlaps are accurately quantified and analysed without redundancy, a method critical for maintaining the integrity of spatial data operations.

 The Jaccard score \( J \) is computed as:
 
\begin{equation}
 J = \frac{\mathrm{TP}}{\mathrm{TP} + \mathrm{FP} + \mathrm{FN}}
\end{equation}

 \( TP \) (True Positives) is the count of precisely positive cases.
 The number of cases mistakenly classified as positive is \( FP \). False Positives.
 The number of positive events wrongly classified as negative is known as \( FN \). False Negatives. The Jaccard score runs from 0 to 1; a score of 1 denotes full agreement between the actual and expected classes, so suggesting no false positives or false negatives.  On the other hand, a score of zero indicates that the expected and actual cases do not coincide. In this work, the quality of the OBD is linked to the degree of overlap, or similarity, and evaluated by assuming OSM is closer to the ground truth. 

%%%%%%%%%%%%%%%%%%%%%%%%%%%%%%%%%%%%%%%%%%
\section{Results}
The following sections discuss the overlap operations performed between OBDs and the OSM building polygons. There are results that are reported in the appendix section for clarity  (see the histograms shown in Figure \ref{fig:building_overlap_hist}). Finally, the log-transformed building sizes are also discussed.

\subsection{Overlap Analysis }
The counts for overlapping polygons should not include any duplicates, and the algorithm, and the data are structured to ensure this. Using unique polygon identifiers (ID), the algorithm computes the number of overlapping OBD and reference polygons, therefore eliminating double counting. The unique count of polygon overlaps is achieved via the Python package, \texttt{pandas} geodataframe's \texttt{nunique()} function. The 'overlapping reference polygons'  and 'overlapping OBD polygons' are synonymous with the concepts of cardinality and ordinality in relational database theory. A polygon can be overlapped by two from another layer (1 to many relationship); conversely, each of the two can be said to be overlapped by one. This approach computes the total count of unique overlaps between the datasets, thereby enabling an exact assessment of spatial overlaps.

\subsubsection{ Microsoft's Building Footprints and OBD buildings}

The overlap analysis between Microsoft's building footprint data and OBD buildings takes stock of the building presence or absence of each dataset relative to the OSM reference dataset across various cities. This section highlights the extent of overlap in building footprints, broadening our understanding of data completeness and accuracy. 

\begin{table}[H]
\centering
\caption{Overlap metrics for overlap between Microsoft data and OSM reference layer. Abbreviations used: Overlapping OBD Polygons (OOP), Overlapping Reference Polygons (ORP), Non-Overlapping OBD Polygons (NOOP), Non-Overlapping Reference Polygons (NORP), OBD with Multiple Overlay (OMO), OBD with Multiple Overlay (RMO), Total number of polygons analysed for the Microsoft data (Total No. OBDs)}

\renewcommand{\arraystretch}{1.3}  % Adjust spacing; 1.3
\begin{tabular}{lccccc}
\toprule
\textbf{Metric} & \textbf{Accra} & \textbf{Caracas} & \textbf{Nairobi} & \textbf{Berlin} & \textbf{Houston} \\
\midrule
Total No. OBDs & \num{1316415} & \num{179303} & \num{325532} & \num{362174} & \num{1363472} \\
Total No. Refs & \num{292038} & \num{12395} & \num{500238} & \num{499530} & \num{791895} \\
OOP (\%)  & \num{10.97} & \num{3.5} & \num{41.29} & \num{62.71} & \num{52.78} \\
ORP (\%) & \num{46.02} & \num{41.03} & \num{25.13} & \num{44.73} & \num{89.39} \\
NOOP (\%)  & \num{89.03} & \num{96.5} & \num{58.71} & \num{37.29} & \num{47.22} \\
NORP (\%) & \num{53.98} & \num{58.97} & \num{74.87} & \num{55.27} & \num{10.61} \\
OMO (\%) & \num{0.0029} & \num{0.0028} & \num{0.03} & \num{0.0144} & \num{0.0044} \\
RMO (\%)  & \num{2.75} & \num{5.82} & \num{1.34} & \num{0.66} & \num{1.37} \\
Avg. IoU Score & 0.5569 & 0.4576 & 0.5165 & 0.6768 & 0.8415 \\
\bottomrule
\end{tabular}
\label{tab:microsoft_overlap_metrics}
\end{table}

Table \ref{tab:microsoft_overlap_metrics} shows results of the overlap analyses for our five study cities. The overlap between Microsoft's building data and the OSM reference layer is computed as described before. Comparing Berlin (62.71\%) and Houston (52.78\%) the calculations results in high proportions of overlapping OBD polygons (OOP). This may indicates a strong alignment between Microsoft and OSM data in those cities. With 41.29\%, Nairobi trailed; Accra (10.97\%) and Caracas (3.5\%) had a comparatively very small degree of overlap.

For Houston, a very high value of 89.39\% was computed, whereas Accra (46.02\%), Berlin (44.73\%), and Caracas (41.03\%), exhibited modest values for overlapping reference polygons (ORP). With 25.13\% Nairobi had the lowest ORP. Non-overlapping OBD polygons (NOOP) were most prevalent in Caracas (96.5\%) and Accra (89.03\%), and least in Berlin (37.29\%).  By contrast, non-overlapping reference polygons (NORP) were lowest in Houston (10.61\%) and greatest in Nairobi (74.87%).

With values below 0.03\% across all cities, Microsoft buildings hardly overlapped many OSM polygons (OMO) regarding multiple overlays. While Berlin had 0.0144\%, Houston had 0.0044\% and Nairobi had the highest OMO (0.03\%). Caracas (5.82\%) had most reference polygons with multiple overlays (RMO), followed by Berlin (0.66\%) and Accra (2.75\%).

The mean IoU values are very high for Houston (0.8415) and Berlin (0.6768), moderate for Accra (0.5569) and Nairobi (0.5165) regions. An IoU value of 0.4576, below 0.5, points to a poor alignment of structures in Caracas. It follows that for cities like Houston and Berlin, Microsoft data might be more in harmony with OSM data, as contrasted with Caracas or Accra.

\subsection {Google Building Footprints and OSM Overlap Analysis }

The overlap analyses of Google's building data with OBD reveal variations in overlap proportions for different metropolitan regions. In this section, the overlap between building footprints and the OSM is described for the two data providers.

\renewcommand{\arraystretch}{1.3}  % Adjust spacing; 1.3 is a good starting point
\begin{table}[H]
\centering
\caption{Overlap Metrics Between Google Data and OSM Reference Layer. Abbreviations used here are the same as those used in Table \ref{tab:microsoft_overlap_metrics}.}
\begin{tabular}{lccc}
\toprule
\textbf{Metric} & \textbf{Accra} & \textbf{Caracas} & \textbf{Nairobi} \\
\midrule
Total OBDs & \num{2619136} & \num{387049} & \num{325532} \\
Total Refs & \num{292038} & \num{12395} & \num{500238} \\
OOP(\%) & \num{8.19} & \num{3.97} & \num{31.63} \\
ORP(\%) & \num{61.03} & \num{57.00} & \num{40.82} \\
NOOP(\%) & \num{91.81} & \num{96.03} & \num{68.37} \\
NORP(\%) & \num{38.97} & \num{43.00} & \num{59.18} \\
OMO(\%) & \num{0.0012} & \num{0.0114} & \num{0.03} \\
RMO(\%) & \num{8.43} & \num{23.58} & \num{7.59} \\
Avg. IoU Score & 0.5086 & 0.2726 & 0.4038 \\
\bottomrule
\end{tabular}
\label{tab:google_overlap_metrics}
\end{table}

Table \ref{tab:google_overlap_metrics} displays intersecting and unique data across cities by aggregating overlap measures between Google OBD and our OBD reference dataset. It emphasises overlapping OBD polygons and overlapping reference polygons. It also lists the total OBD and total reference polygons for Google and OBD data.

Over the three cities studied, Google building data and the OSM reference layer had different overlapping metrics. While the percentage of overlapping reference polygons (ORP) was 61.03\%, 8.19\% of the Google building polygons in Accra overlapped OSM polygons (OOP). On the other hand, in Caracas, only 3.97\% of the Google polygons overlapped (OOP). However 57.00\% of the OSM reference polygons matched (ORP). From the Google perspective, Nairobi displayed a really significant overlap: 31.63\% OOP but 40.82\% ORP.

Concerning the non-overlapping cases, the number of unmatched Google polygons (NOOP) was highest in Caracas (96.03\%), followed by Accra (91.81\%) and Nairobi (68.37\%). Nairobi got the highest value (59.18\%) followed by Accra and Caracas with 38.97\% and 43.00\% correspondingly in terms of unmatched reference polygons (NORP).

Accra had a minimal proportion of Google polygons with multiple overlays (OMO = 0.0012), while Caracas and Nairobi recorded somewhat higher values of 0.0114 and 0.03, respectively. For polygons with many overlaps, Accra had a negligible fraction of Google polygons. For reference polygons with several overlays (RMO),  8.43\% were found in Accra, 23.58\% in Caracas, and 7.59\% in Nairobi. 

Reflecting the spatial alignment of matched polygons, the average IoU scores for Accra, Caracas, and Nairobi were 0.5086, 0.2726, and 0.4038, respectively. These results imply rather low agreement in Caracas but rather modest spatial alignment in Accra and Nairobi.
\subsection{Percentile Analysis of Microsoft OBD Building Sizes}

By examining both log-transformed (see Figure \ref{fig:log_poly_pdf} for the normal curve fitted on the log-transformed distribution) and original data values, building sizes in square metres, the dataset's spread and skewness are visible. The  10th, 25th, 50th (median), 75th, and 90th percentiles of building sizes (areas) for Accra, Caracas, Houston, Nairobi, and Berlin are represented. Data on the Microsoft OBD in this particular analysis to understand the trend in size distribution across the study cities is reported. Building sizes tend to be smaller in global South areas where 'kiosk estates' or 'kiosk compounds' exist \citep{alba2022}; these are makeshift dwellings usually found in urban informal settlements.

\begin{table}[H]
    \centering
    \caption{Percentiles of original building sizes. This is based on the original data without a transformation.}
    \begin{tabular}{lccccc}
        \toprule
        Percentiles & Accra & Caracas & Houston & Nairobi & Berlin \\
        \midrule
        10th & 27.06 & 28.66 & 86.13 & 30.60 & 25.70 \\
        25th & 45.27 & 43.85 & 152.46 & 52.67 & 43.57 \\
        50th & 101.22 & 73.86 & 216.52 & 111.34 & 109.42 \\
        75th & 194.02 & 136.44 & 284.56 & 220.45 & 186.75 \\
        90th & 292.41 & 284.13 & 415.17 & 405.35 & 445.38 \\
        \bottomrule
    \end{tabular}
    \label{tab:percentiles_original}
\end{table}

Table \ref{tab:percentiles_original} presents the percentiles for the original building sizes in each city. These values provide a clear picture of the individual building size distribution and the central tendency without transformation. The variability observed across different cities underscores the uniqueness in data representation.

%%normal pdf fitted

\subsection{Log-Transformed Building Polygon Size Distributions}

The following figures illustrate the probability density functions (PDF) of the log-transformed polygon sizes for different cities. The objective of all transformations is to generate a relatively symmetric distribution.  This establishes a solid foundation for further statistical methodologies.  The transformed distribution does not have to be entirely normal, but if it is, this would enhance confidence in tests utilising smaller samples and could streamline statistical models \citep{West2022}. Each figure contains a histogram representing the empirical distribution, for Microsoft OBD and a fitted normal curve. The datasets originate from Google and Microsoft sources, as indicated in the captions.

\begin{figure}[H]
    \centering
    \begin{tabular}{cc}
        \parbox{0.45\textwidth}{
            \centering
            \includegraphics[width=0.45\textwidth]{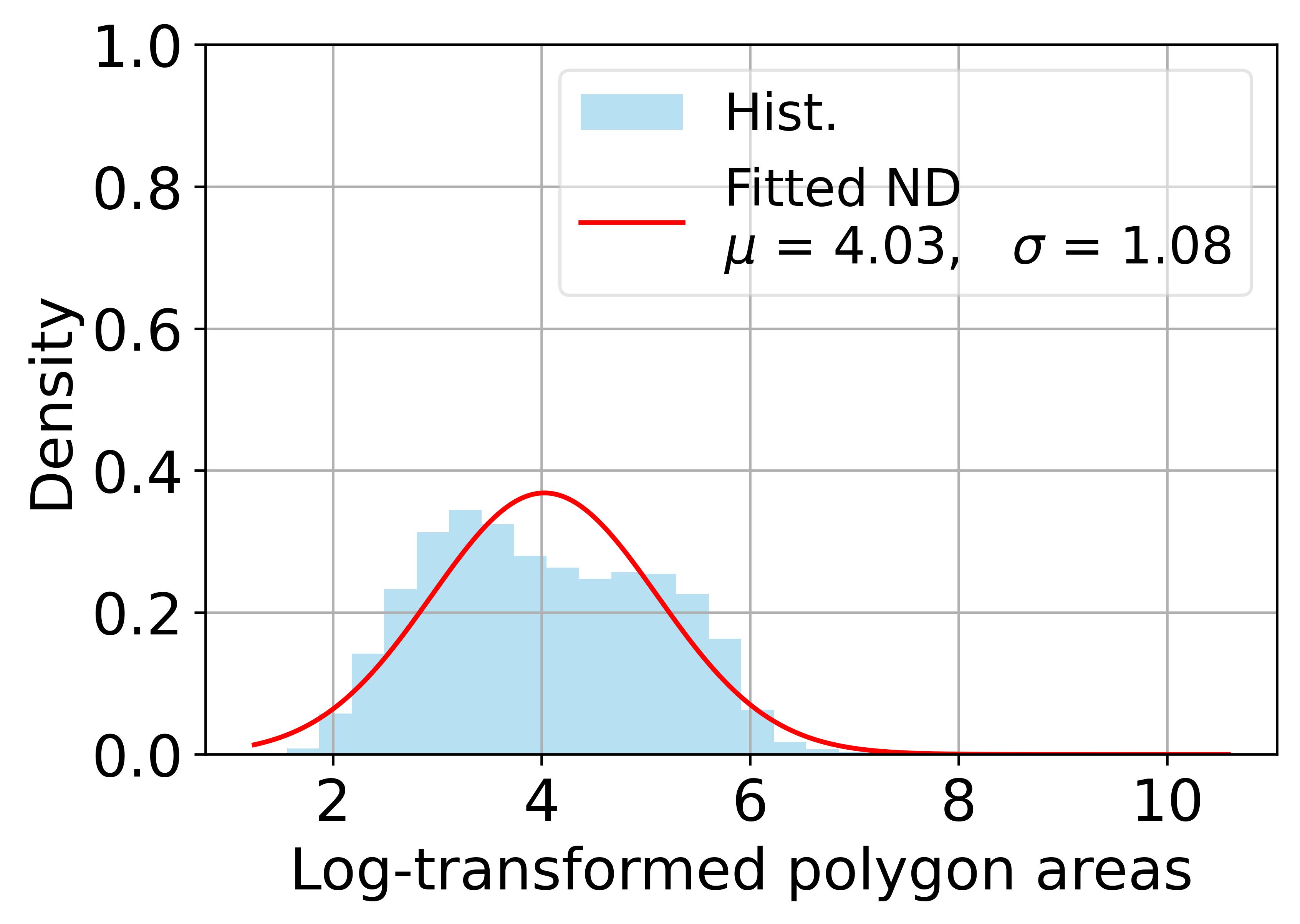} \\
            (a) Accra - Google
        }
        &
        \parbox{0.45\textwidth}{
            \centering
            \includegraphics[width=0.45\textwidth]{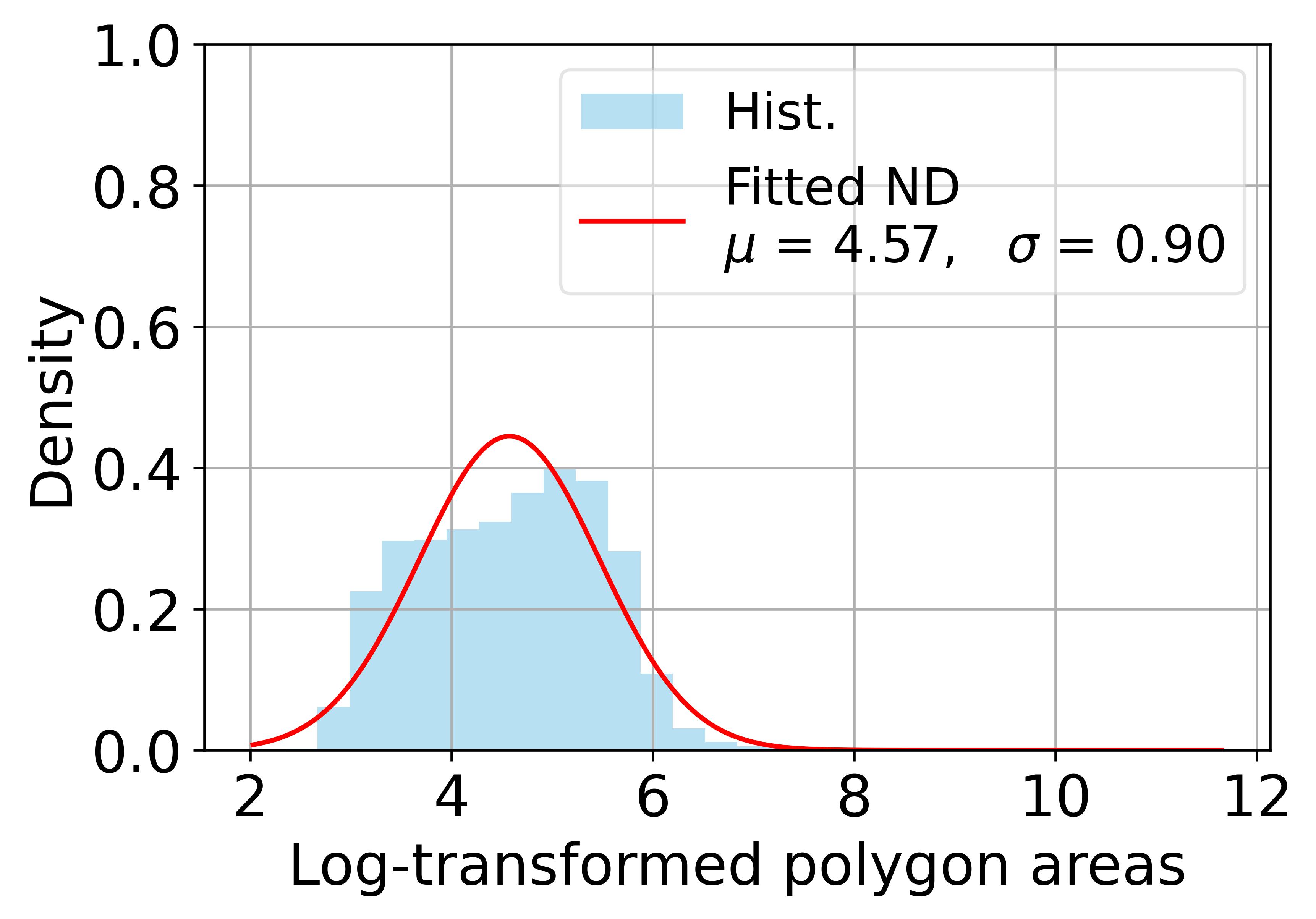} \\
            (b) Accra - Microsoft
        }\\
        \parbox{0.45\textwidth}{
            \centering
            \includegraphics[width=0.45\textwidth]{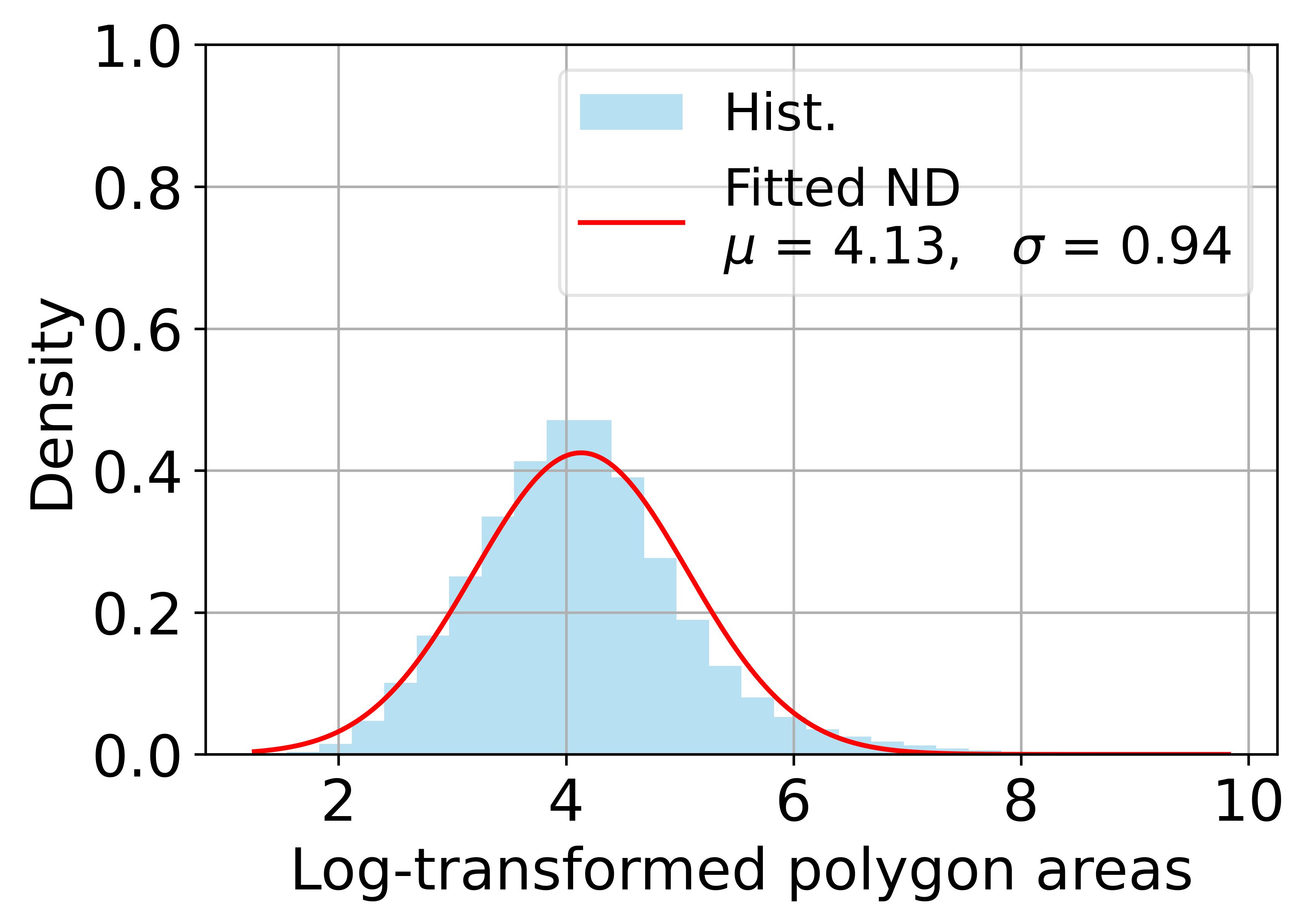} \\
            (c) Caracas – Google
        }
        &
        \parbox{0.45\textwidth}{
            \centering
                \includegraphics[width=0.45\textwidth]{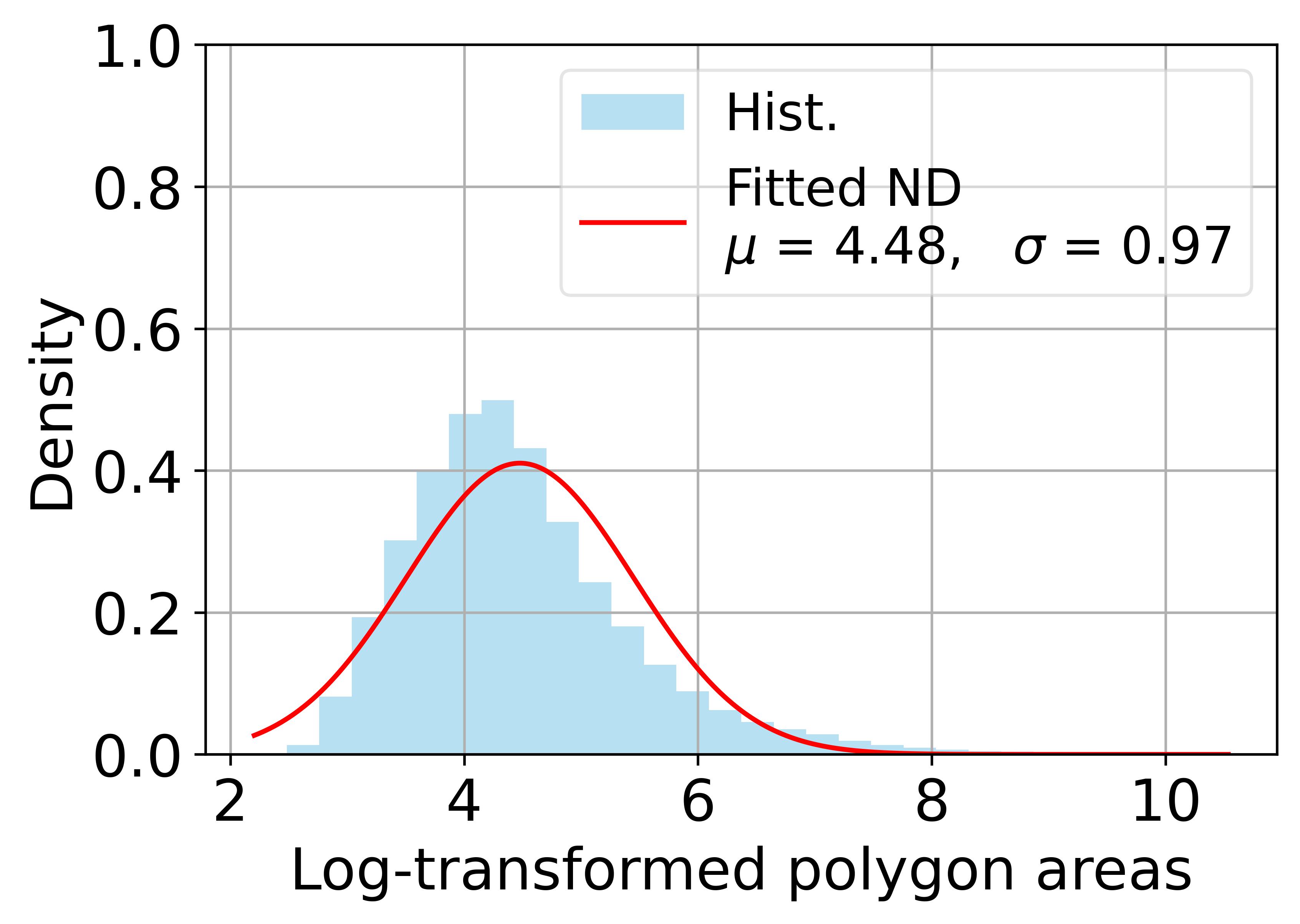} \\
            (d) Caracas - Microsoft
        }\\
        \parbox{0.45\textwidth}{
            \centering
            \includegraphics[width=0.45\textwidth]{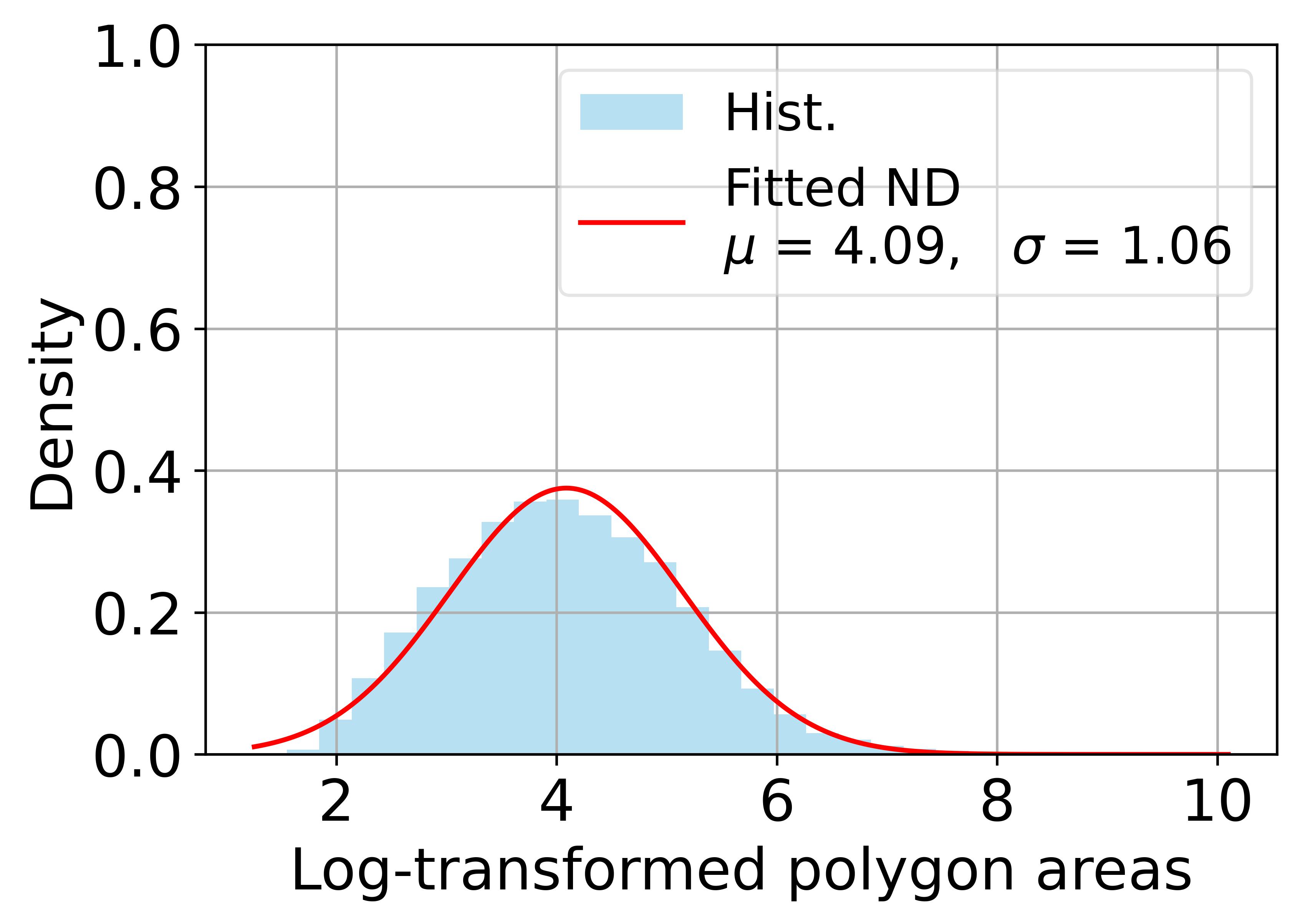} \\
            (e) Nairobi – Google
        }
        &
        \parbox{0.45\textwidth}{
            \centering
            \includegraphics[width=0.45\textwidth]{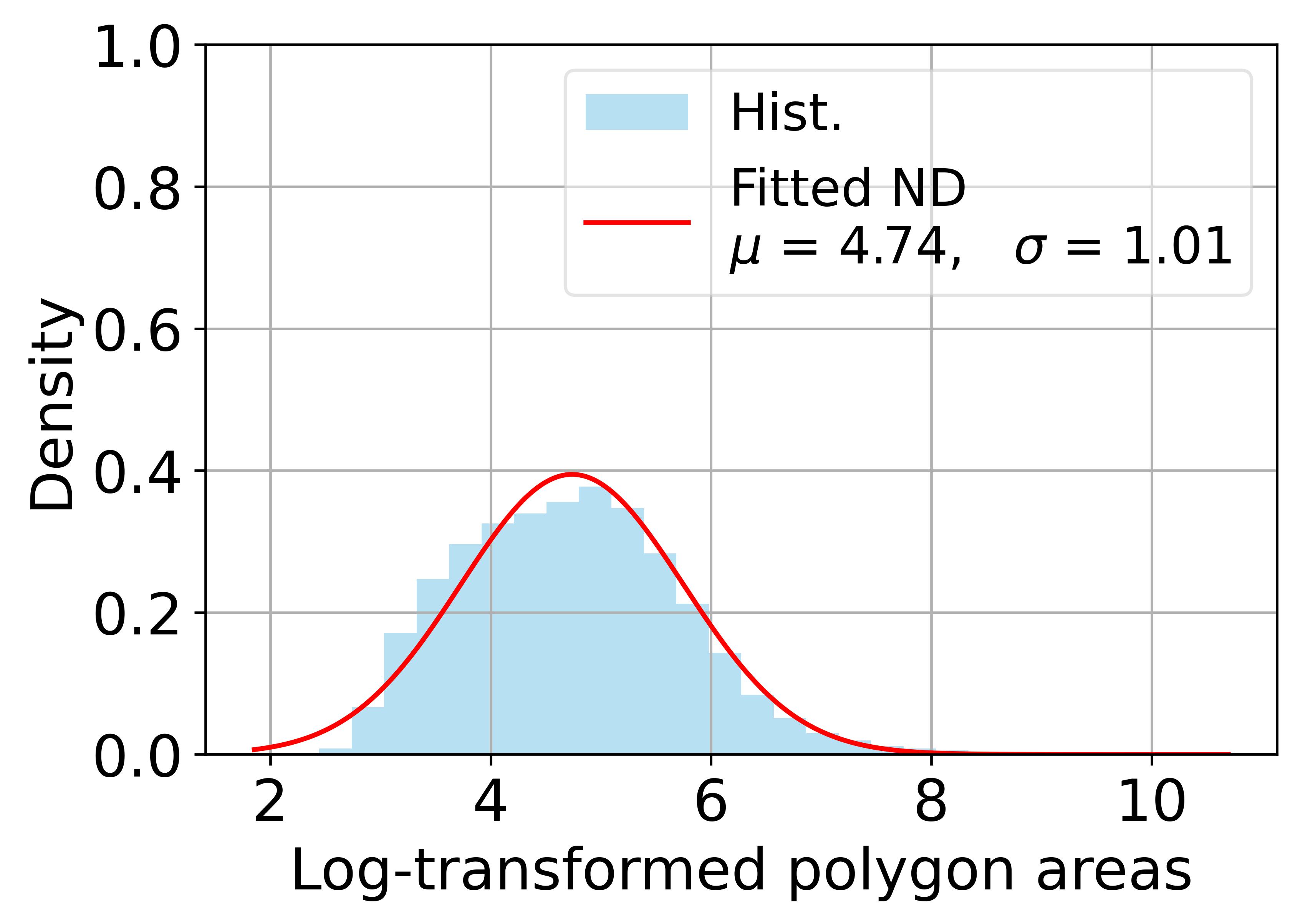} \\
            (f) Nairobi - Microsoft
        }\\
        \parbox{0.45\textwidth}{
            \centering
            \includegraphics[width=0.45\textwidth]{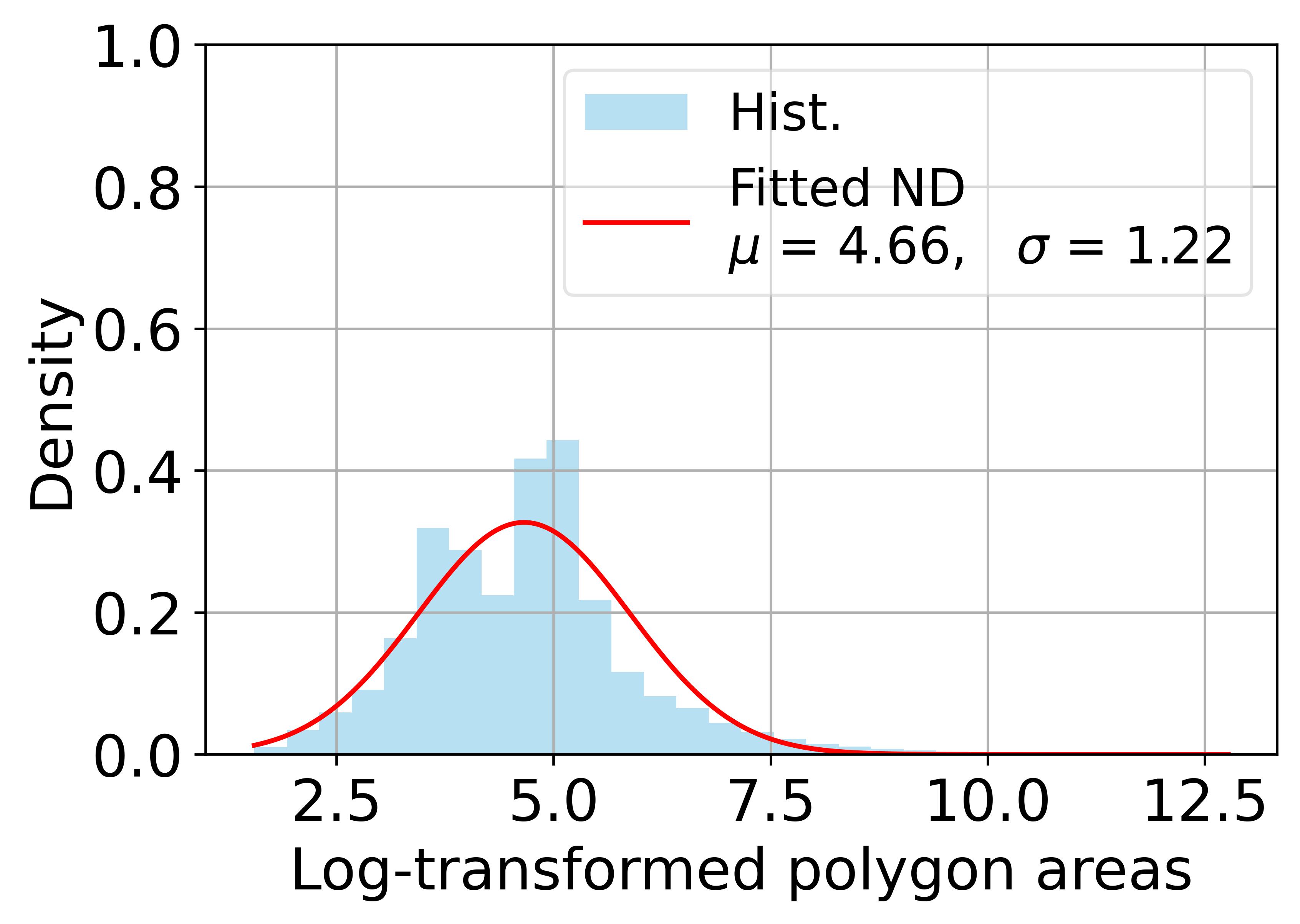} \\
            (g) Berlin
        }
        &
        \parbox{0.45\textwidth}{
            \centering
            \includegraphics[width=0.45\textwidth]{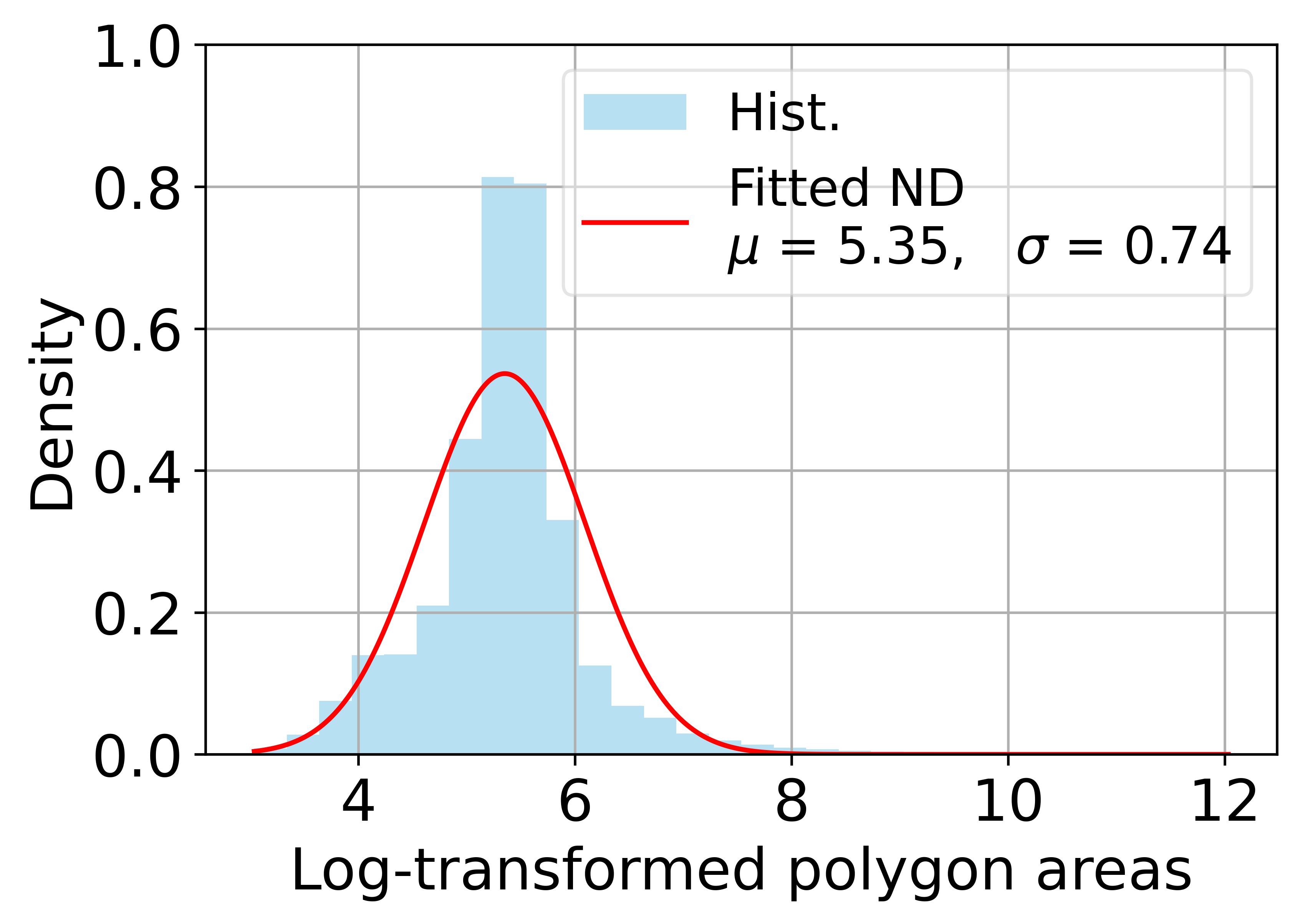} \\
            (h) Houston
        }
    \end{tabular}
    \caption{Log-transformed building polygon size distributions for selected cities. Each panel represents the empirical histogram and fitted normal distribution.}
    \label{fig:log_poly_pdf}
\end{figure}

The distribution of building sizes in a wide range of urban settings is compared in Table \ref{fig:log_poly_pdf}. Inference can be made by observing the log-transformed and original data percentiles in detail. The mean inferred from the PDF of the log-transformed data is useful for understanding typical building sizes in deprived areas; smaller values are recorded for the Google OBDs rather than the Microsoft OBD. The highest and lowest mean building sizes are computed for Houston and Accra, respectively. Areas with economic activity tend to have larger structures, representing factory outlets, huge shopping malls and generally larger houses and flats.

\subsection{Matching of Homologous Points Across Datasets}
In this subsection, a comparison of geometric measurements among buildings in five diverse cities is made. These geometric values are computed using the two datasets and cross-referenced against OSM building data. The key variables analysed include the mean differences of inner angles (in decimal degrees) and incoming and outgoing bearings (measured in a clockwise fashion with respect to the True North). The distances, or displacements, between homologous points which represent the vertices of building polygons in metres (as detailed in Table~\ref{tab:buildings_vertices_comparison}) are computed in a pairwise manner.

For the Google dataset, the data indicates the following intuitions:

\begin{itemize}
    \item In Accra, the mean angle difference is approximately 0.30°, with the incoming and outgoing bearings both recorded at about 1.77°. The distance between homologous points is evaluated at 1.30 m.
    
    \item Caracas exhibits a mean angle difference of 0.44°, with bearings moderately adjusted from 1.44° to 1.41°, and a homologous point distance (HPD) of 1.25 m.

    \item Nairobi presents a mean angle difference of 0.40° and bearing adjustments ranging from 1.82° to 1.85°, with a corresponding HPD of 1.26 m.
    
\end{itemize}

In contrast, the Microsoft dataset displays more pronounced angle differences: Accra's mean angle difference is recorded at 1.23°, slightly exceeding those from the Google dataset, with bearings marginally fluctuating around 1.94° and a distance of 1.25 m. Next is Caracas, with a mean angle difference of 1.35°, with bearings from 2.02° to 2.00°, and an increased homologous point distance of 1.38 m. Nairobi follows with an angle difference of 1.42°, which is the highest amongst all cities in both datasets, with bearings marginally consistent around 1.93° and a point distance of 1.17 m. The Data for Berlin reveals a mean angle difference of 0.43°, with bearings recorded at approximately 1.87° and a distance of 1.17 m. While, in Houston, the lowest mean angle difference of 0.23° is observed, with bearings near 0.69° and a distance of 1.52 m noted.

As outlined in Table~\ref{tab:buildings_vertices_comparison}, these observations suggest variance in data acquisition methods, resolution, or representation between the Google and Microsoft datasets relative to the OBD reference. Microsoft generally presents larger angle differences than Google for analogous city data, indicating potential discrepancies in the geometric accuracy of building structures.

\begin{table}[H]
  \centering
  \caption{Comparison of Matched Points for Building Datasets across Cities. Distances are in metres and angles in decimal degrees. Angle represents the mean inner clockwise angle formed at all vertices. Bearing1 (Brg1) is the clockwise bearing, moving in the clockwise direction, from the previous vertex to any vertex. Bearing2 (Brg2) is the clockwise bearing, moving in a clockwise fashion, from any vertex to the next vertex. Distance (Dist) is the Euclidean distance between a point in the layer being compared and the reference layer. }
  \label{tab:buildings_vertices_comparison}
  \begin{tabular}{cccccc}
    \toprule
    Provider & City & Angle (deg) & Brg1 (deg) & Brg2 (deg) & Dist. (m) \\
    \midrule
    Google & Accra & 0.3017 & 1.7653 & 1.7729 & 1.2993 \\
    & Caracas & 0.4431 & 1.4434 & 1.4131 & 1.2547 \\
    & Nairobi & 0.3963 & 1.8224 & 1.8497 & 1.2631 \\
    
    \midrule
    Microsoft & Accra & 1.2323 & 1.9422 & 1.9433 & 1.2489 \\
    & Caracas & 1.3463 & 2.0218 & 1.9994 & 1.3785 \\
    & Nairobi & 1.4164 & 1.9302 & 1.9330 & 1.1709 \\
    & Berlin & 0.4271 & 1.8699 & 1.8739 & 1.1702 \\
    & Houston & 0.2423 & 0.7152 & 0.7130 & 1.3069 \\
    \bottomrule
  \end{tabular}
\end{table} 

The tables above indicate various discrepancies between the two datasets across the measured metrics, highlighting how each city's building morphology might influence the measurements.

\subsection{Computation of Intersection over Union }

The following figures present the IoU or Jaccard similarity index values obtained for building footprints across different cities. Then again, in this work, significance filtering guarantees that the IoU is computed considering only significant overlaps between polygons, constituting matching criteria. Setting a criterion such that only intersections covering at least 51\% of an OBD polygon are regarded as important helps the analysis avoid skewing results with minor or edge overlaps without considerable harmony between datasets. To emphasise, the algorithm computes iou for all relevant intersections, avoiding double computation. 

The IOU statistics are visualised using the \texttt{Viridis} colour scale, where variations indicate differences in the spatial accuracy of the extracted footprints. Each city is represented in two columns: the left column corresponds to Google data, and the right column corresponds to Microsoft data; Berlin data is an exception and uses a hybrid dataset.

\begin{figure}[H]
    \centering
    \renewcommand{\arraystretch}{1.2}
    \begin{tabular}{c c}
        \parbox{0.45\textwidth}{\includegraphics[width=0.45\textwidth]{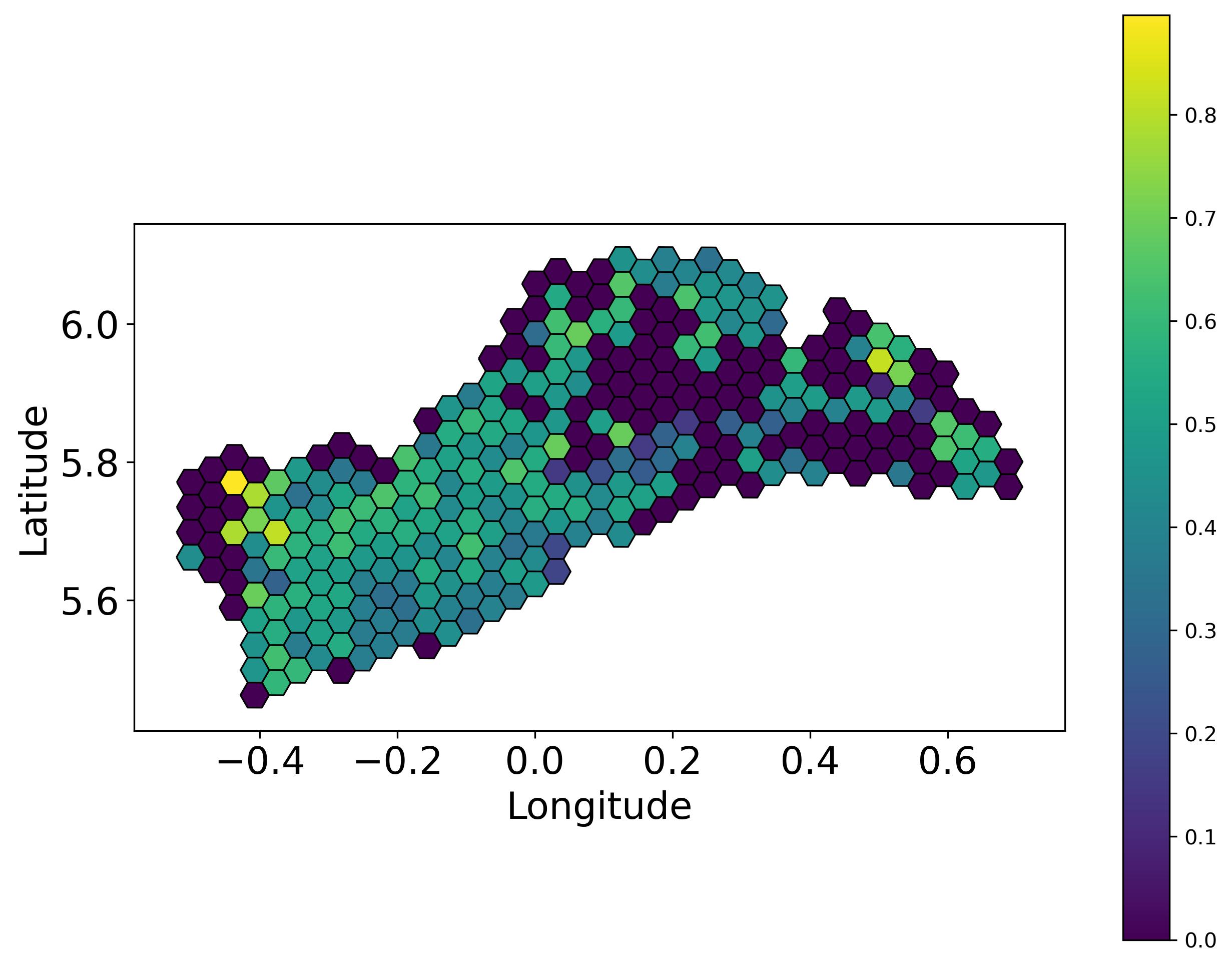} \\ (a) Google Data - Accra} &
        \parbox{0.45\textwidth}{\includegraphics[width=0.45\textwidth]{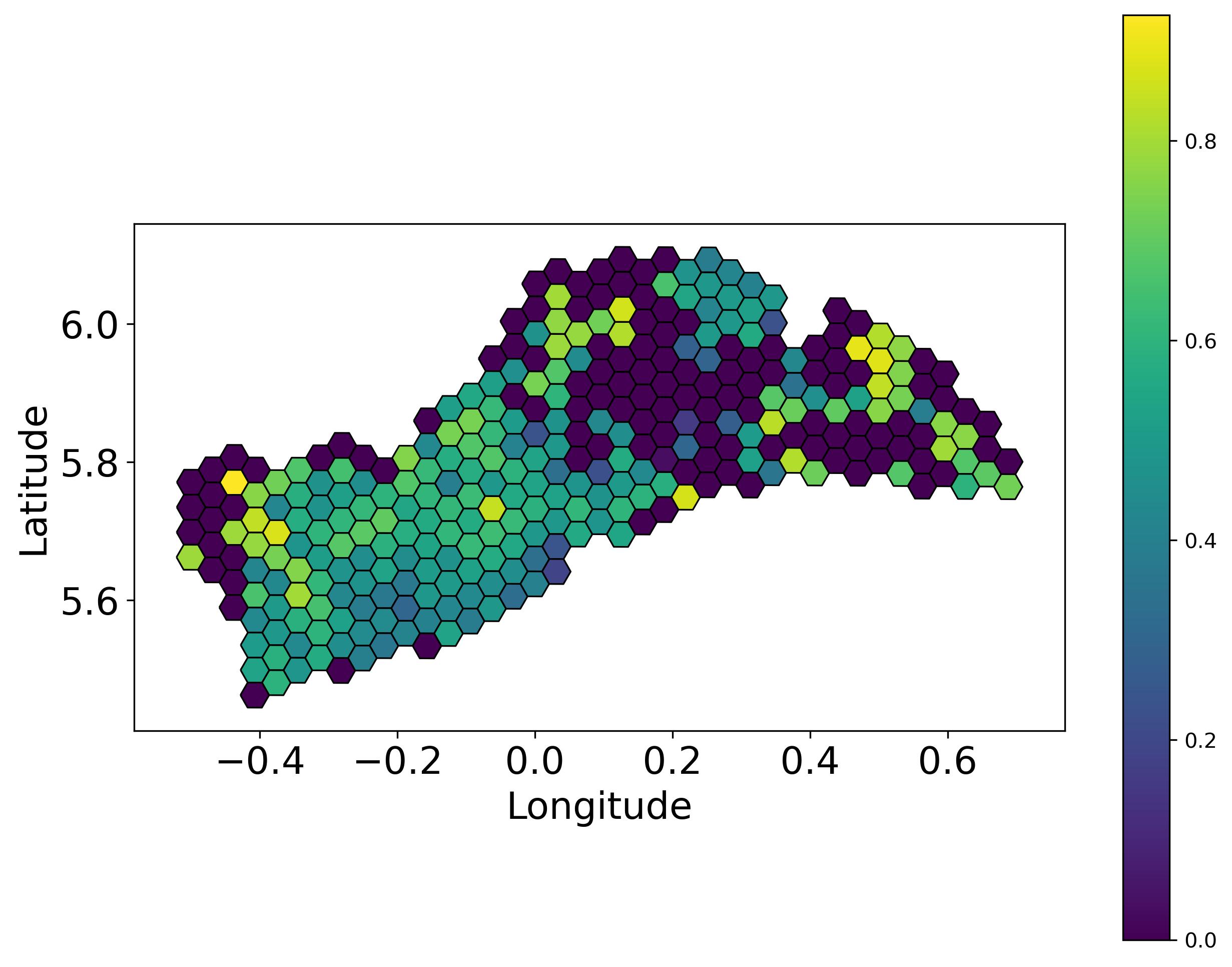} \\ (b) Microsoft Data - Accra} \\
        \parbox{0.45\textwidth}{\includegraphics[width=0.45\textwidth]{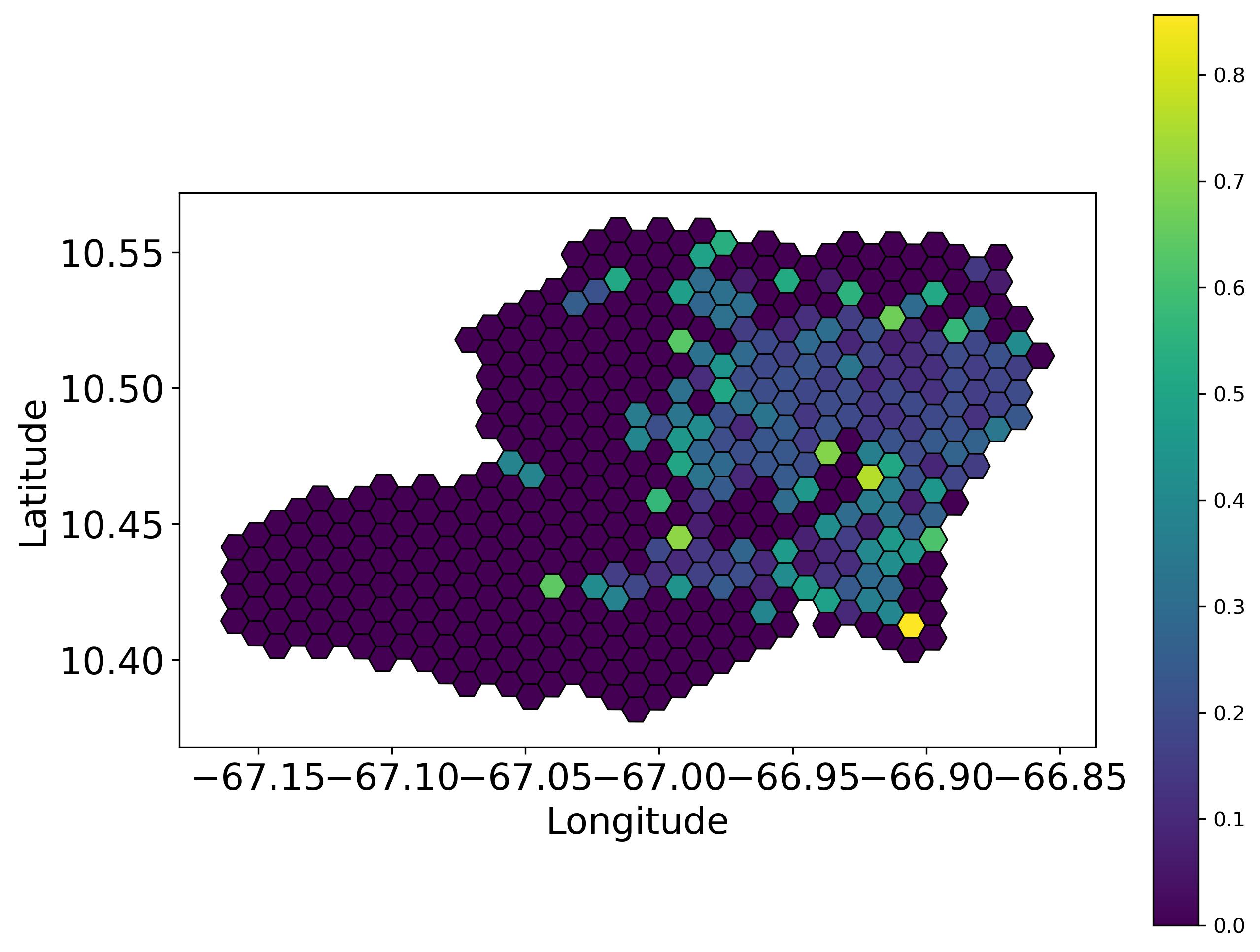} \\ (c) Google Data - Caracas} &
        \parbox{0.45\textwidth}{\includegraphics[width=0.45\textwidth]{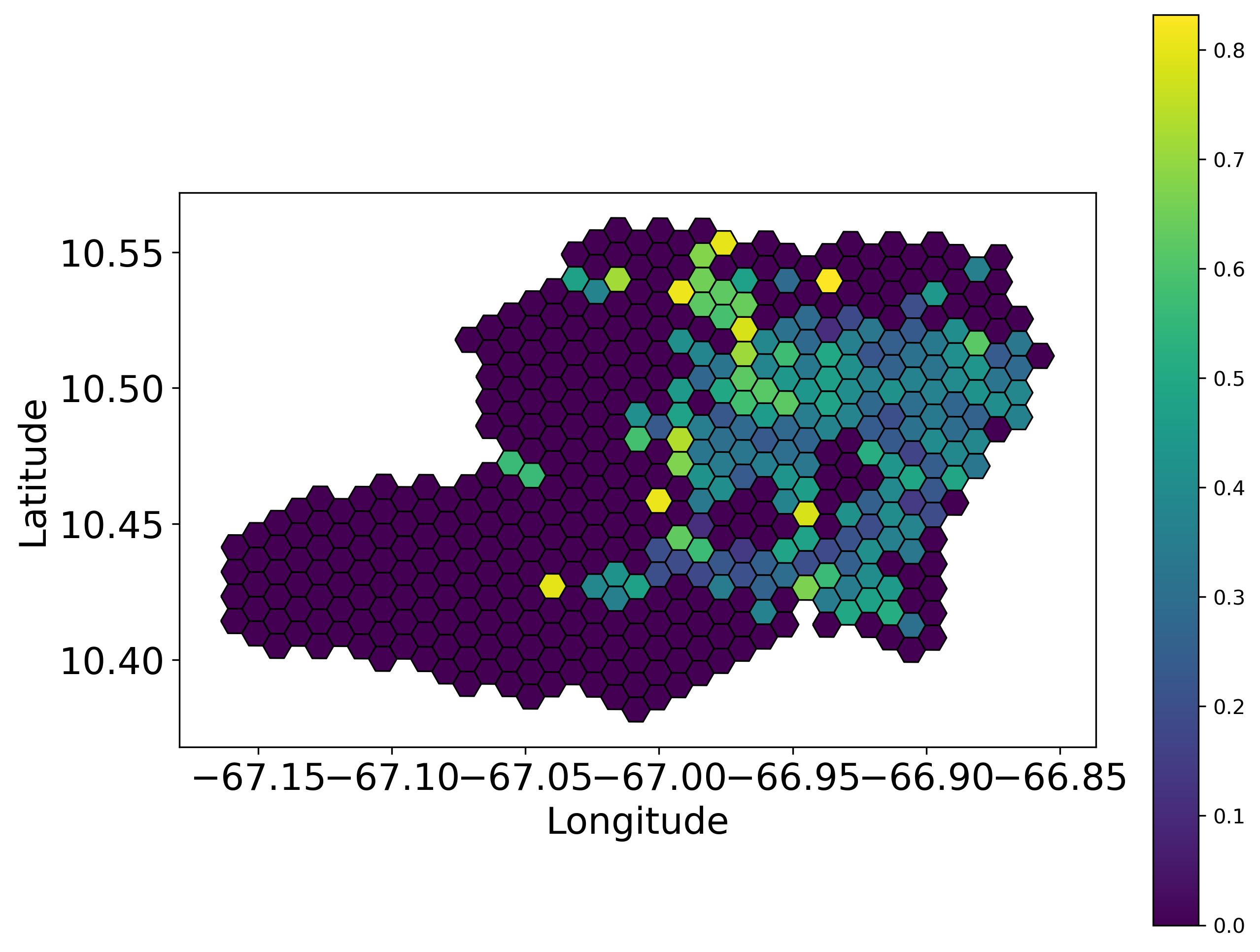} \\ (d) Microsoft Data - Caracas} \\
        \parbox{0.45\textwidth}{\includegraphics[width=0.45\textwidth]{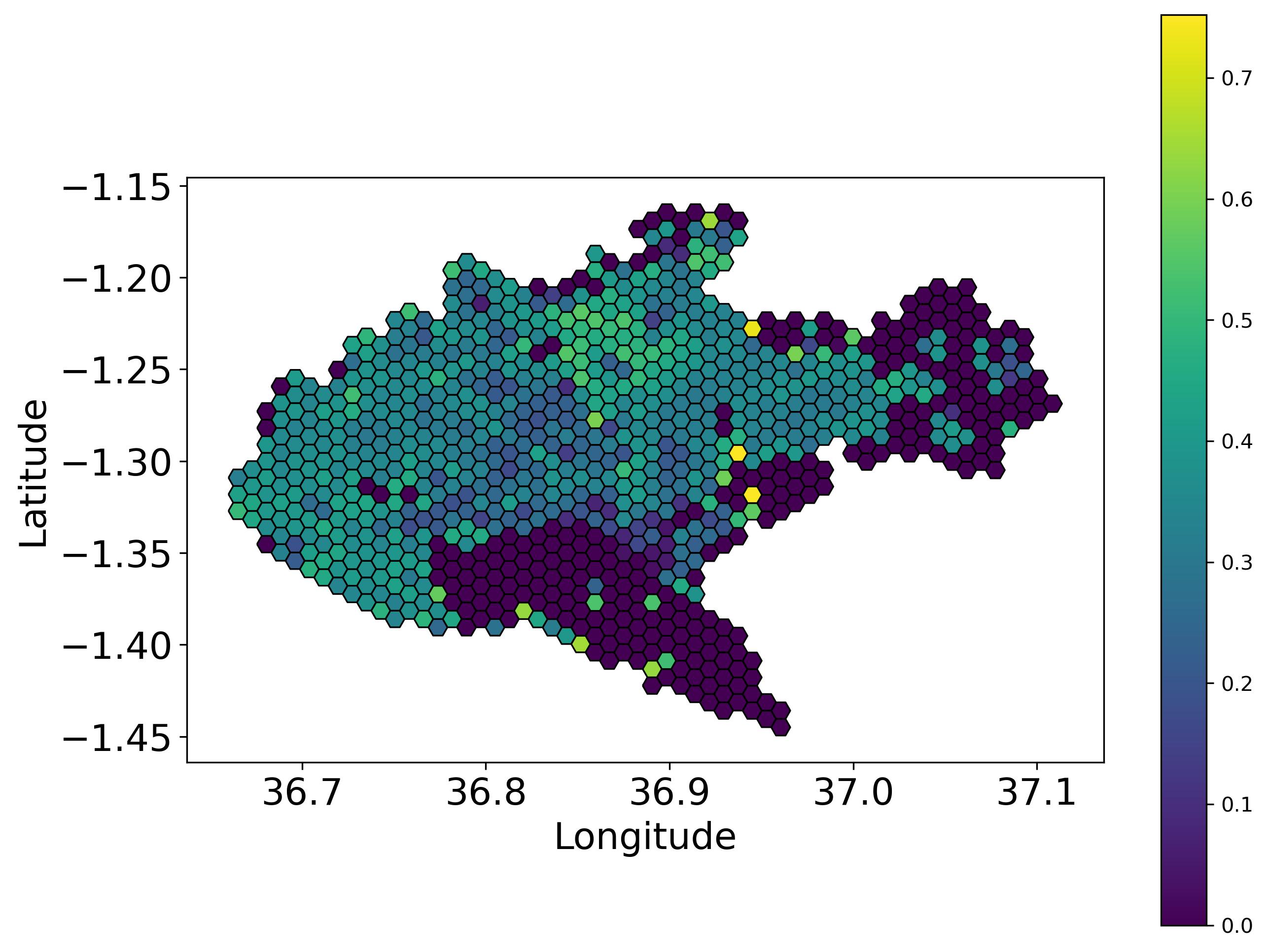} \\ (e) Google Data - Nairobi} &
        \parbox{0.45\textwidth}{\includegraphics[width=0.45\textwidth]{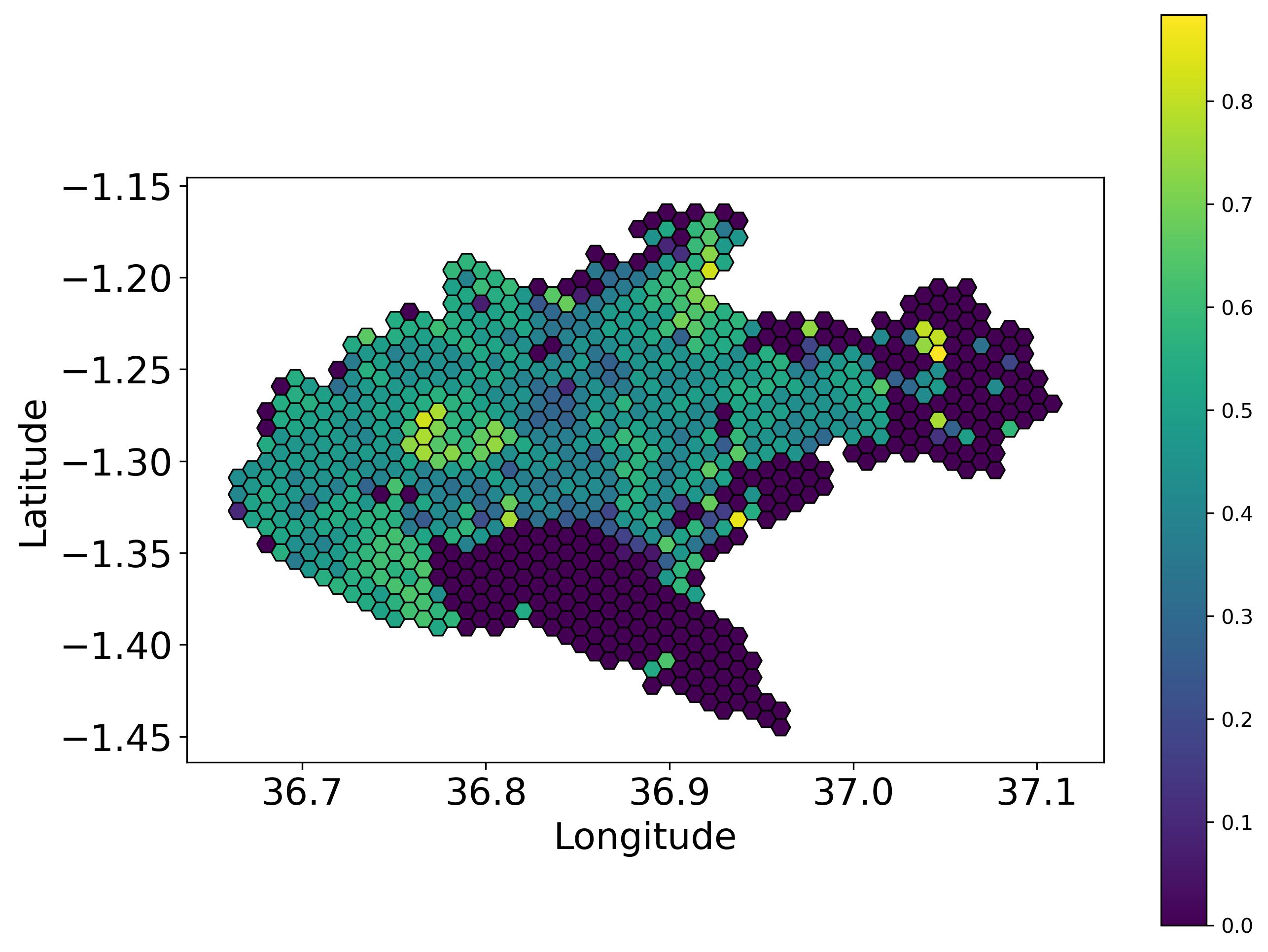} \\ (f) Microsoft Data - Nairobi} \\
        \parbox{0.45\textwidth}{\includegraphics[width=0.45\textwidth]{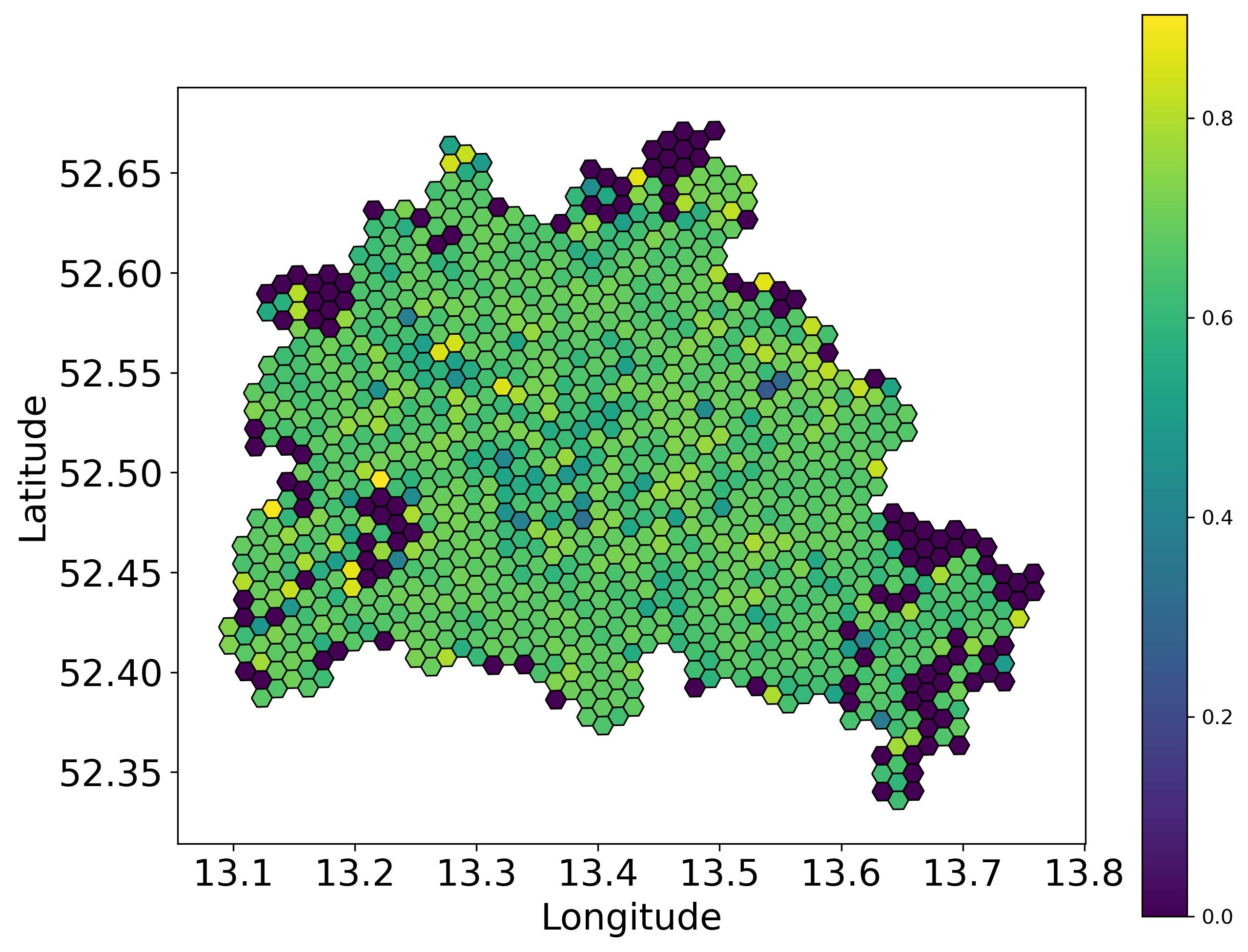} \\ (g) VIDA Data - Berlin} &
        \parbox{0.45\textwidth}{\includegraphics[width=0.45\textwidth]{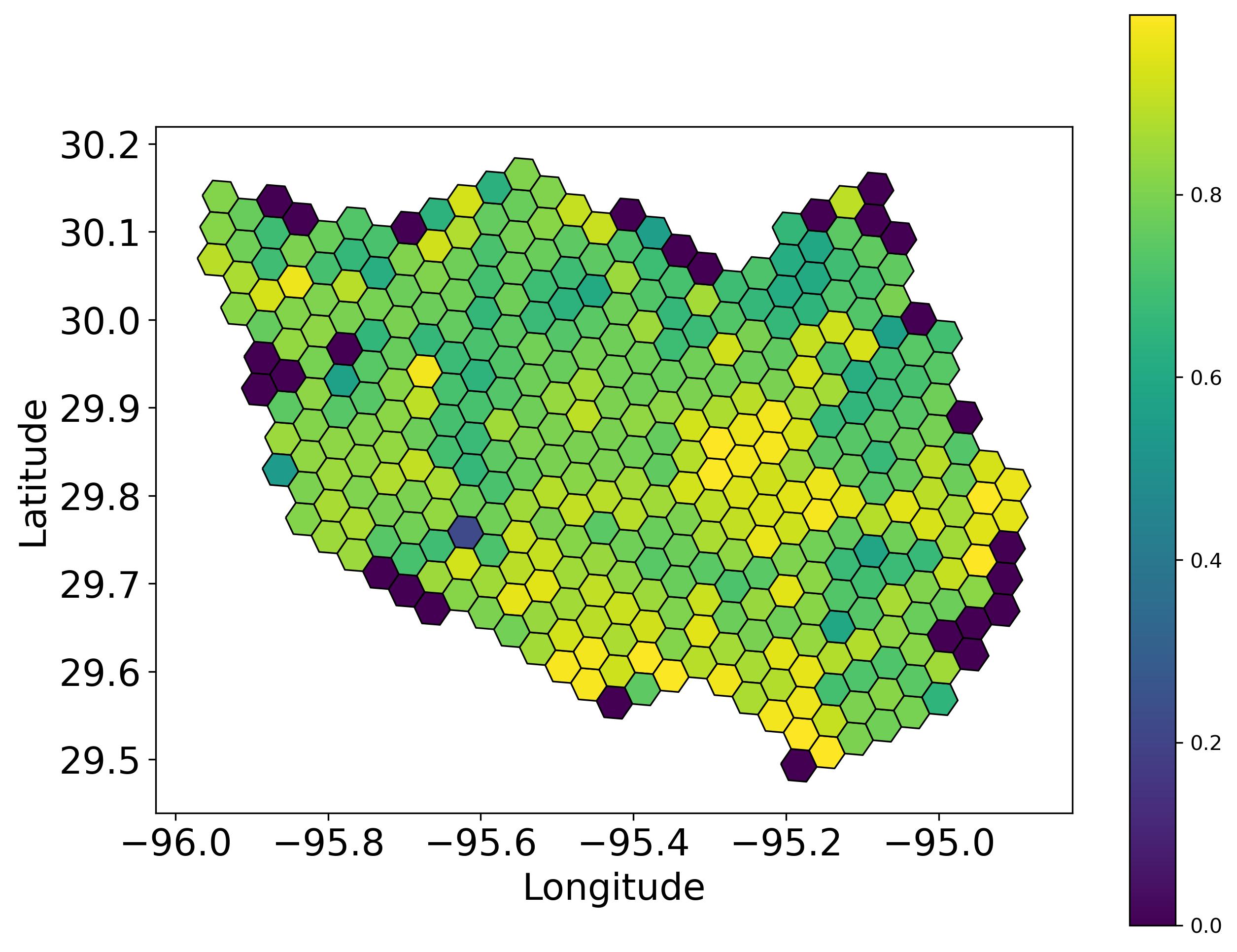} \\ (h) Microsoft Data – Houston}
    \end{tabular}
    \caption{A visualisation of IoU values for building footprints across cities. Each row represents a city, with the left column depicting Google data and the right column depicting Microsoft data. The exception is Berlin, which uses a hybrid dataset, Microsoft's VIDA. The colour scale \texttt{Viridis} is used to indicate spatial agreement, with higher IoU values representing better alignment with reference data.}
    \label{fig:iou_comparison_map}
\end{figure}

Using building footprint statistics from Google, Microsoft, and VIDA (for Berlin only), the hexbin maps in Figure \ref{fig:iou_comparison_map} displays the spatial distribution of IoU for the study cities. With isolated areas with significant concurrence in the northeastern and central areas of Accra, both Google (a) and Microsoft (b) data show usually low to moderate accuracies. Whereas the Microsoft dataset (d) shows more generally moderate values, notably in the northern region, suggesting a rather improved spatial alignment, the Google dataset (c) shows mostly low values with limited central clustering for Caracas.

With Google (e) and Microsoft (f), whose data shows somewhat more high-value hexagons, Nairobi shows more overall spatial coverage and moderately higher IoU values for both providers. With a uniform distribution of moderate to high values and limited low-value areas, Berlin's VIDA dataset (g) exhibits regularly high IoU values throughout the city. At last, the Microsoft dataset for Houston (h) shows a substantial concentration of high within---cell similarity, particularly in the central and eastern zones. This emphasises good agreement or data coverage. 

%%%%%%%%%%%%%%%%%%%%%%%%%%%%%%%%%%%%%%%%%%
\section{Discussion}
In the following sections an examination of the IoU score is described. The evaluation of spatial overlaps in urban building datasets using the average IoU scores is discussed. The assumption of higher conformity in structured cities like Houston and Berlin can be justified by inconsistencies in spatial representation for cities such as Accra and Nairobi. The study demonstrates directional and linear displacements, therefore showing difficulties in accurately depicting building structures in complex urban settings—especially in the global South. Although the study offers information on data accuracy, its reliance on certain datasets and measurements opens up possible biases, implying the need for further studies.

 Starting off with the Jaccard or IoU score metric, the average IoU scores for Accra and Nairobi is 0.5086 and 0.4038, respectively. This may suggest rather larger degrees of overlap, implying rather a degree of spatial inconsistency or a chaotic scene. In comparison, the relative accuracy of the Caracas OBD with respect to the OSM data informs the low IoU value of 0.2726 together with a high proportion of non-overlapping OBD polygons (96.03\%). The high values of overlapping reference polygons (ORP) in Accra (61.03\%) and Caracas (57.00\%) suggest that Google's feature extraction might have under-represented or overlooked several buildings even if reference buildings were covered.  With average IoU scores of 0.8415 and 0.6768, respectively, Microsoft's building data on the other hand, showed greater conformity with OSM in cities like Houston and Berlin. The accuracy of Microsoft's collection in metropolitan environments with structured layouts. These cities also noted significant overlap percentages for both OBD and reference polygons. On the contrary, Caracas once more showed low agreement (IoU = 0.4576), in line with the Google results, implying possible difficulties in capturing building structures in more uneven urban fabric; this challenge is also attributed to complex morphology \citep{Kraff2020a}.

While similarity scores may be helpful as a quality metric the relative positional accuracy can confirm the quality assessment. If there is bias in data quality between cities. Tables \ref{tab:buildings_vertices_comparison} and \ref{tab:percentiles_original} show accuracy variations and polygon size diversity in cities, respectively. Percentiles and log-transformed size distributions, seen in Figure \ref{fig:log_poly_pdf}, indicate significant changes in spatial representation variability. This could highlight data collection biases or urban heterogeneity at the building level or the existence of both phenomena. The mean building size follows an interesting trend, with cities in the global South having lower values. 

Building footprint overlaps and alignment errors between datasets are used to infer positional and directional accuracies across cities. Tables \ref{tab:google_overlap_metrics} and \ref{tab:microsoft_overlap_metrics} show the proportion of overlapping and non-overlapping polygons, while the means of the relative positions of polygon vertices of the layers are compared in Table \ref{tab:buildings_vertices_comparison}. For example Nairobi has the worst directional error when it comes to the alignment of the building with respect to OSM, while Houston buildings only deviate by a mean of 0.6 degrees. Contrarily, the directional error for Nairobi, though not the best, is very close to that of Berlin, in the global North. It follows that while mean metrics indicate relative positional accuracy, they also highlight differences in angle, bearing, and distance. This could possibly point to diversity in data acquisition, modelling and post-processing methods necessary. 

The study's focus on five locations constrains its ability to adequately depict worldwide urban diversity. Dependency on OSM as a reference resource implies its correctness, which varies over different regions. Also our work assesses Google and Microsoft's AI-generated datasets. Even though they are easily accessible and truly open, this excludes other pertinent data sources. The study makes use of particular quality measures which might not cover all facets of data quality, such as temporal or thematic accuracy. Especially in informal communities, acknowledged biases in AI models and the presumption of OSM as a ground truth could influence assessments.  The above limitations may suggest areas for further research, while our findings contribute to the quality assessment of OBDs.

%%%%%%%%%%%%%%%%%%%%%%%%%%%%%%%%%%%%%%%%%%
%\section{Conclusions}
This study contributes to research on the evaluation of the reliability of automatically extracted open building datasets (OBDs). This is an increasingly critical data source for urban planning, mapping, and monitoring deprived regions. A systematic comparison of Google and Microsoft OBDs to OSM reference data across diverse global cities has been carried out. The analyses reveal significant spatial variation in data quality, particularly in cities characterised by complex morphologies. These findings confirm that AI-generated building footprints can vary in accuracy and completeness across geographic contexts. It also raises important questions about potential biases in data acquisition and modelling practices.

The importance of this study lies in three aspects. Firstly, it engages with questions of data equity, which is important as data providers have focused on certain parts of the globe. Secondly, it focuses on representational accuracy, which is an important basis for analysis at all possible scales. Finally, the practical limitations of using AI in urban planning are revealed through our quantitative approach. Open data and machine learning are being increasingly integrated into official mapping efforts and urban analytics, and our findings urge caution; stakeholders should not rely on such datasets without rigorous quality assessment. By proposing a quantitative, comparative approach to evaluating OBDs, this research offers a framework that can be applied in future studies and extended to additional cities. The work makes more evident the necessity of ground-truthed validation. It also stresses transparency in modelling practices and attention to urban context when adopting AI-derived datasets for decision-making in the geospatial domain. 
% This section is not mandatory, but can be added to the manuscript if the discussion is unusually long or complex.

%%%%%%%%%%%%%%%%%%%%%%%%%%%%%%%%%%%%%%%%%%
% \section{Patents}

% This section is not mandatory, but may be added if there are patents resulting from the work reported in this manuscript.

%%%%%%%%%%%%%%%%%%%%%%%%%%%%%%%%%%%%%%%%%%
\vspace{6pt} 

%%%%%%%%%%%%%%%%%%%%%%%%%%%%%%%%%%%%%%%%%%
%% optional
%\supplementary{The following supporting information can be downloaded at:  \linksupplementary{s1}, Figure S1: title; Table S1: title; Video S1: title.}

% Only for journal Methods and Protocols:
% If you wish to submit a video article, please do so with any other supplementary material.
% \supplementary{The following supporting information can be downloaded at: \linksupplementary{s1}, Figure S1: title; Table S1: title; Video S1: title. A supporting video article is available at doi: link.}

% Only used for preprtints:
% \supplementary{The following supporting information can be downloaded at the website of this paper posted on \href{https://www.preprints.org/}{Preprints.org}.}

% Only for journal Hardware:
% If you wish to submit a video article, please do so with any other supplementary material.
% \supplementary{The following supporting information can be downloaded at: \linksupplementary{s1}, Figure S1: title; Table S1: title; Video S1: title.\vspace{6pt}\\
%\begin{tabularx}{\textwidth}{lll}
%\toprule
%\textbf{Name} & \textbf{Type} & \textbf{Description} \\
%\midrule
%S1 & Python script (.py) & Script of python source code used in XX \\
%S2 & Text (.txt) & Script of modelling code used to make Figure X \\
%S3 & Text (.txt) & Raw data from experiment X \\
%S4 & Video (.mp4) & Video demonstrating the hardware in use \\
%... & ... & ... \\
%\bottomrule
%\end{tabularx}
%}

%%%%%%%%%%%%%%%%%%%%%%%%%%%%%%%%%%%%%%%%%%

\section*{Abbreviations}
\begin{tabular}{ll}
OBD & Open Building Dataset\\
OSM & OpenStreetMap\\
VIDA & Visualization and Interactive Data Analysis\\
\end{tabular}

%%%%%%%%%%%%%%%%%%%%%%%%%%%%%%%%%%%%%%%%%%
\appendix
%\section[\appendixname~\thesection]{} 
\section{overlap analysis}
\label{sec:overlap-analysis}

\begin{figure}[H]
    \centering
    \renewcommand{\arraystretch}{1.2}
    \begin{tabular}{c c}
        \parbox{0.40\textwidth}{\includegraphics[width=0.40\textwidth]{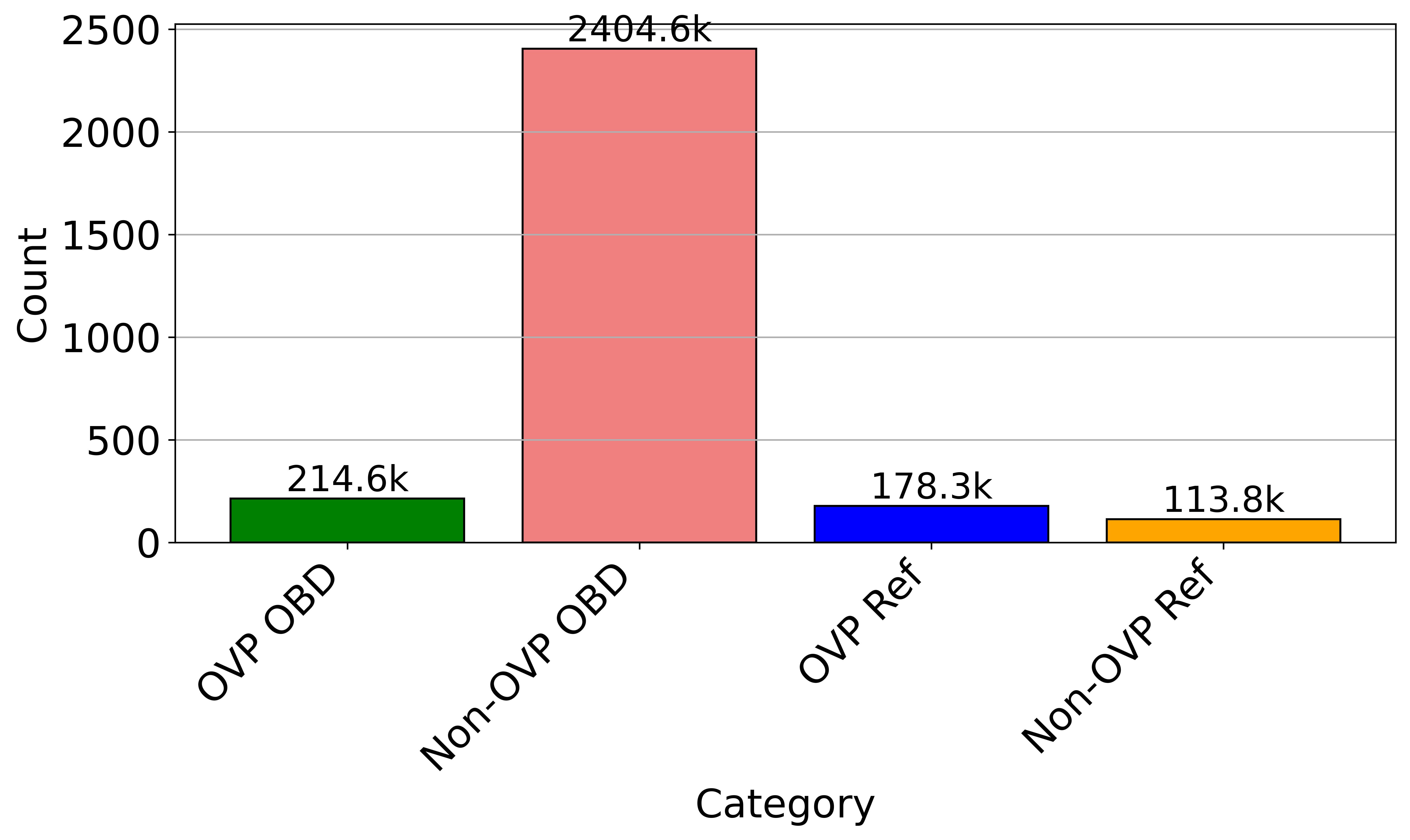} \\ (a) Accra --- Google } &
        \parbox{0.40\textwidth}{\includegraphics[width=0.40\textwidth]{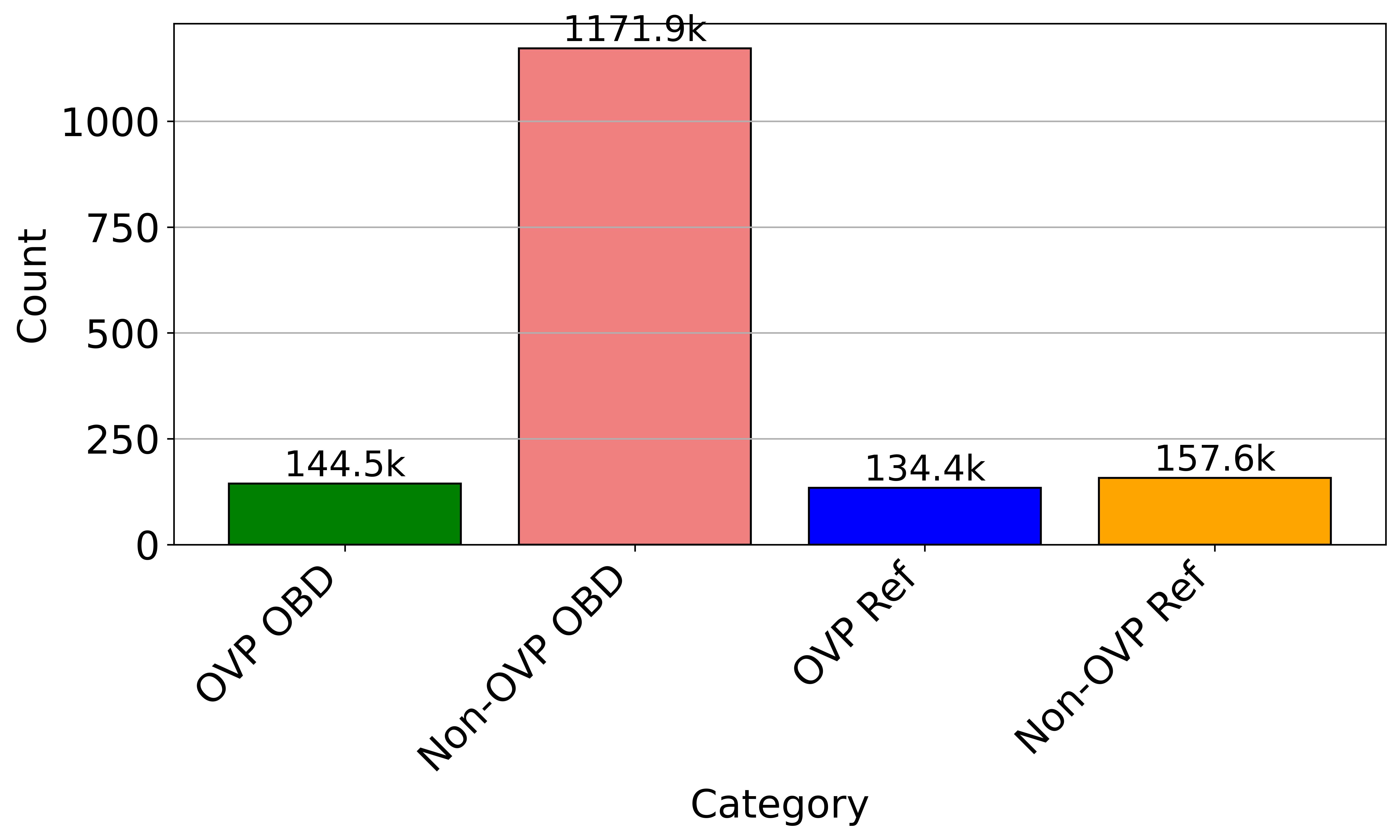} \\ (b) Accra --- Microsoft } \\
        \parbox{0.40\textwidth}{\includegraphics[width=0.40\textwidth]{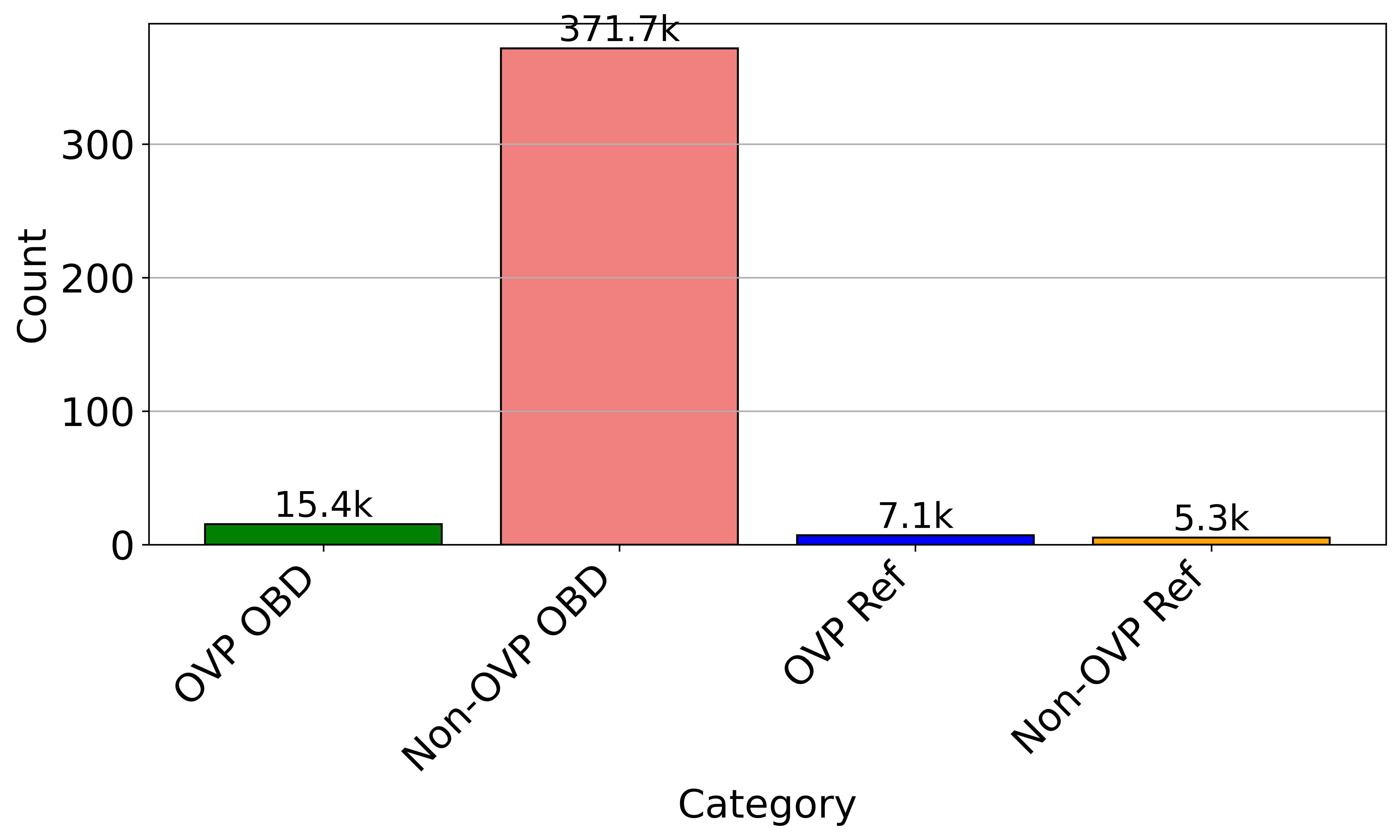} \\ (c) Caracas --- Google } &
        \parbox{0.40\textwidth}{\includegraphics[width=0.40\textwidth]{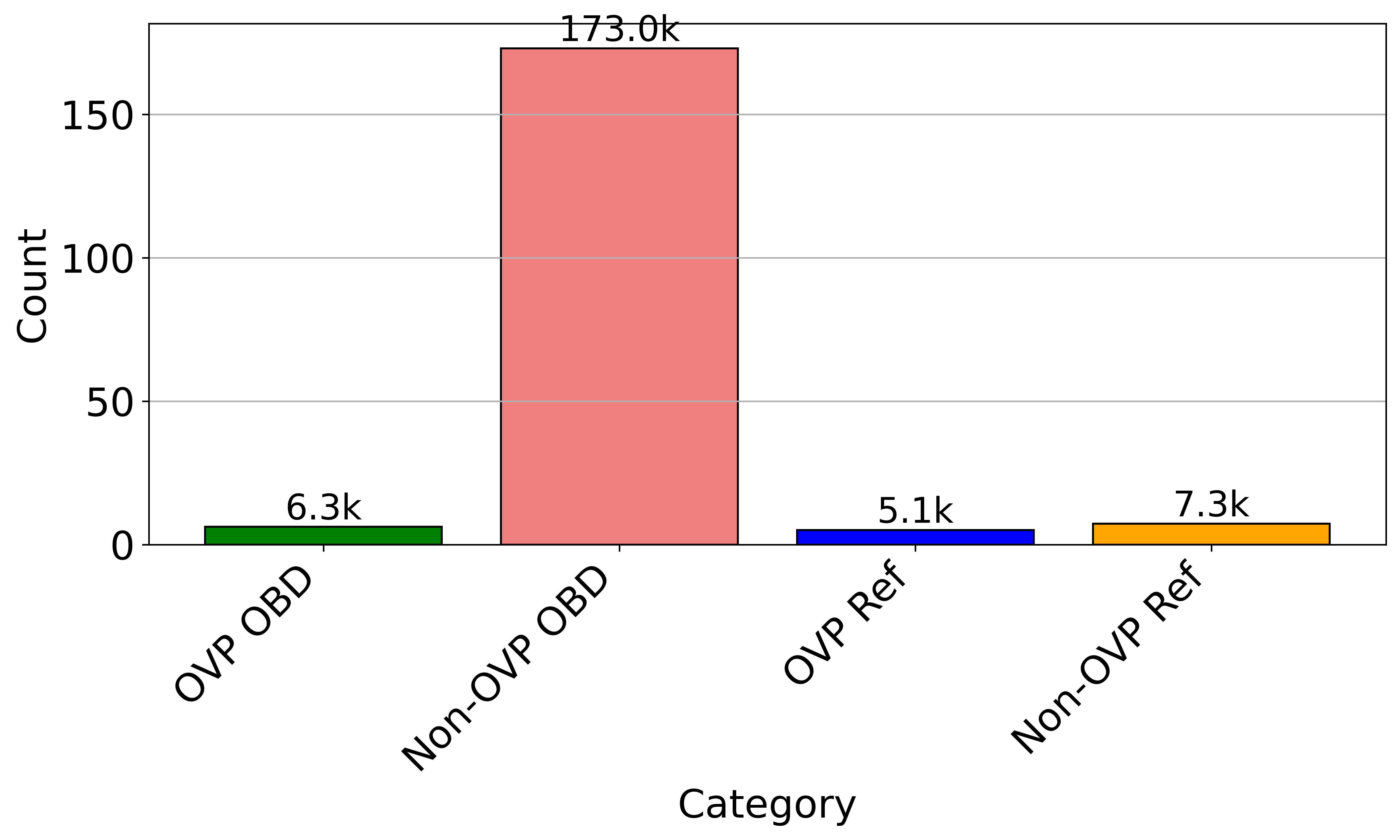} \\ (d) Caracas --- Microsoft } \\
        \parbox{0.40\textwidth}{\includegraphics[width=0.40\textwidth]{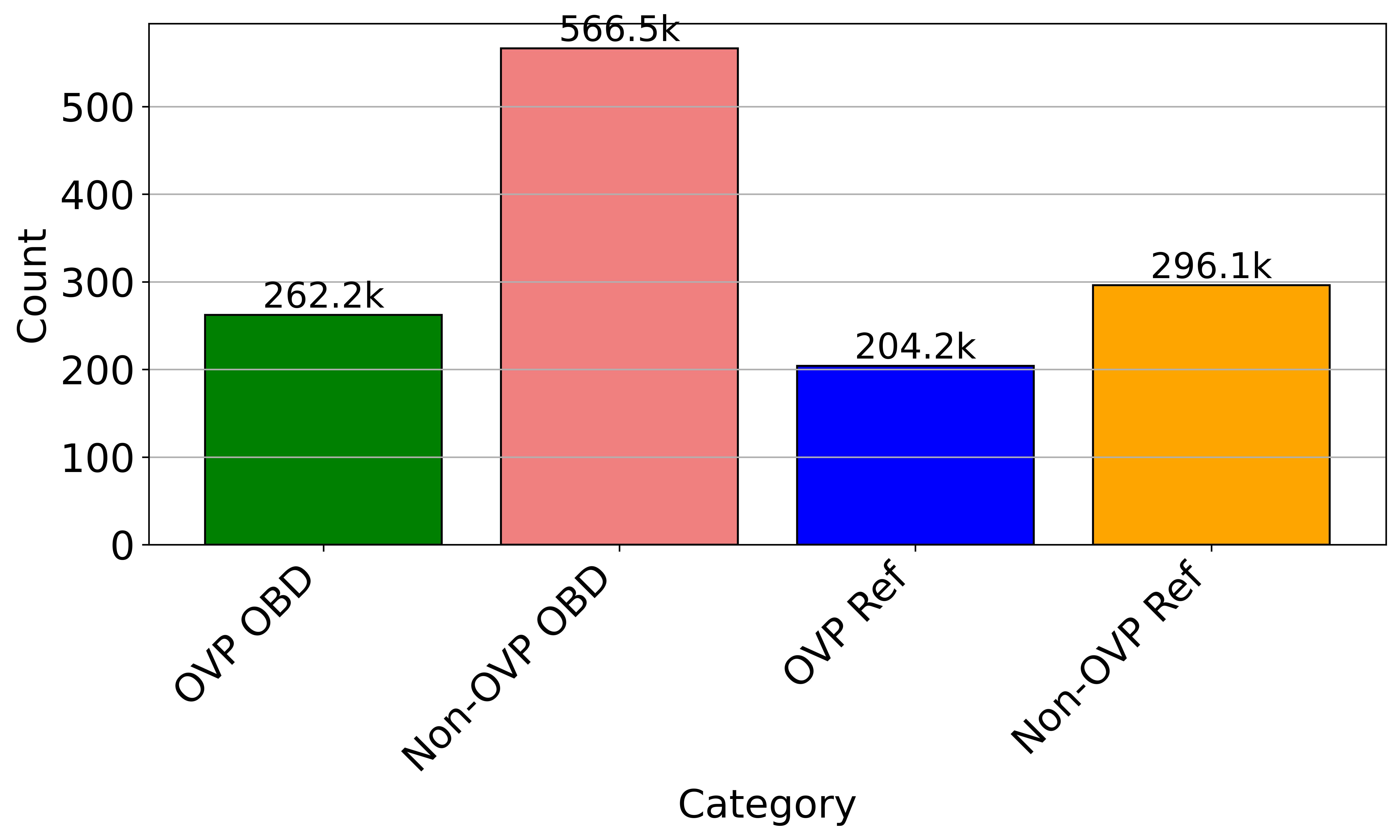} \\ (e) Nairobi --- Google } &
        \parbox{0.40\textwidth}{\includegraphics[width=0.40\textwidth]{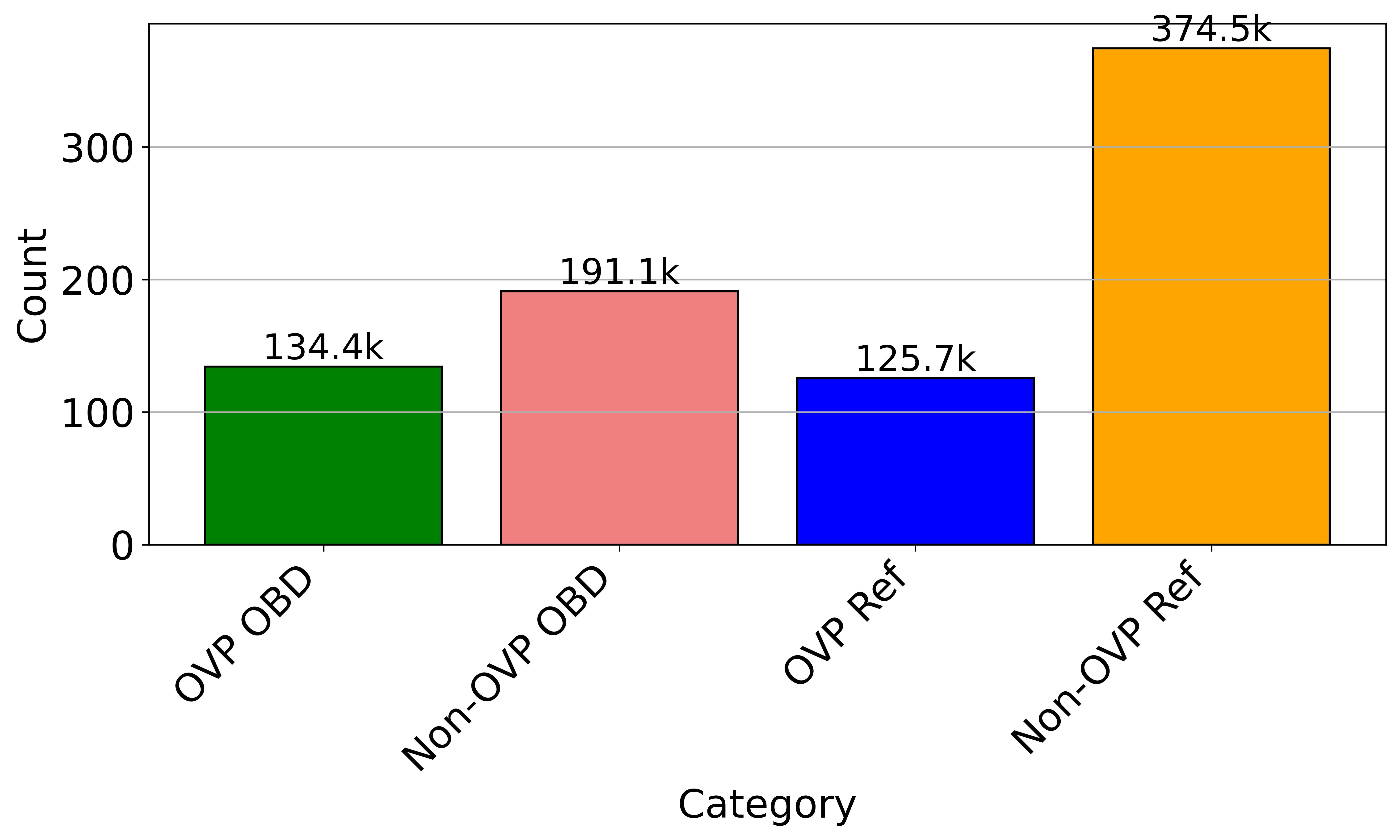} \\ (f) Nairobi --- Microsoft } \\
        \parbox{0.40\textwidth}{\includegraphics[width=0.40\textwidth]{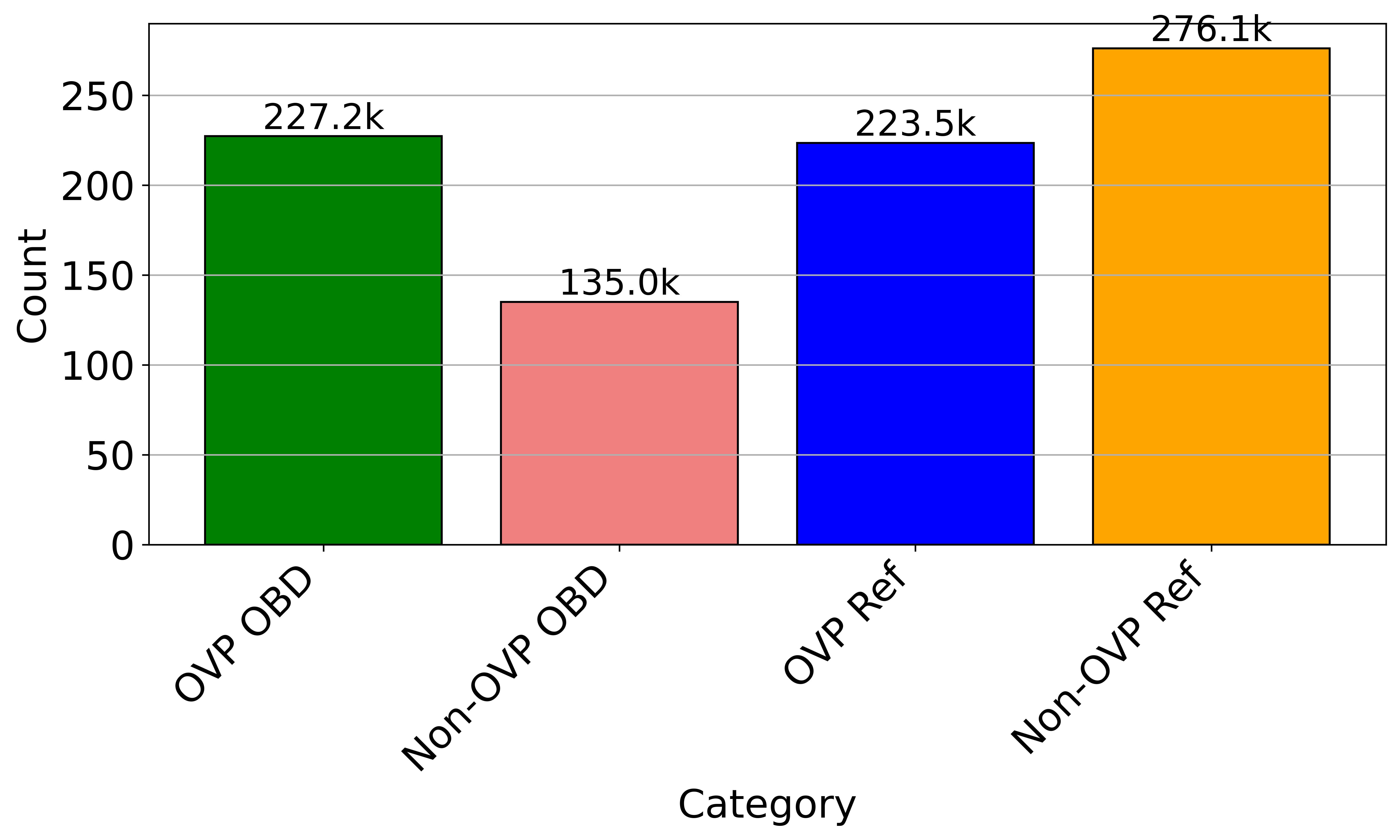} \\ (g) Berlin --- Microsoft VIDA } &
        \parbox{0.40\textwidth}{\includegraphics[width=0.40\textwidth]{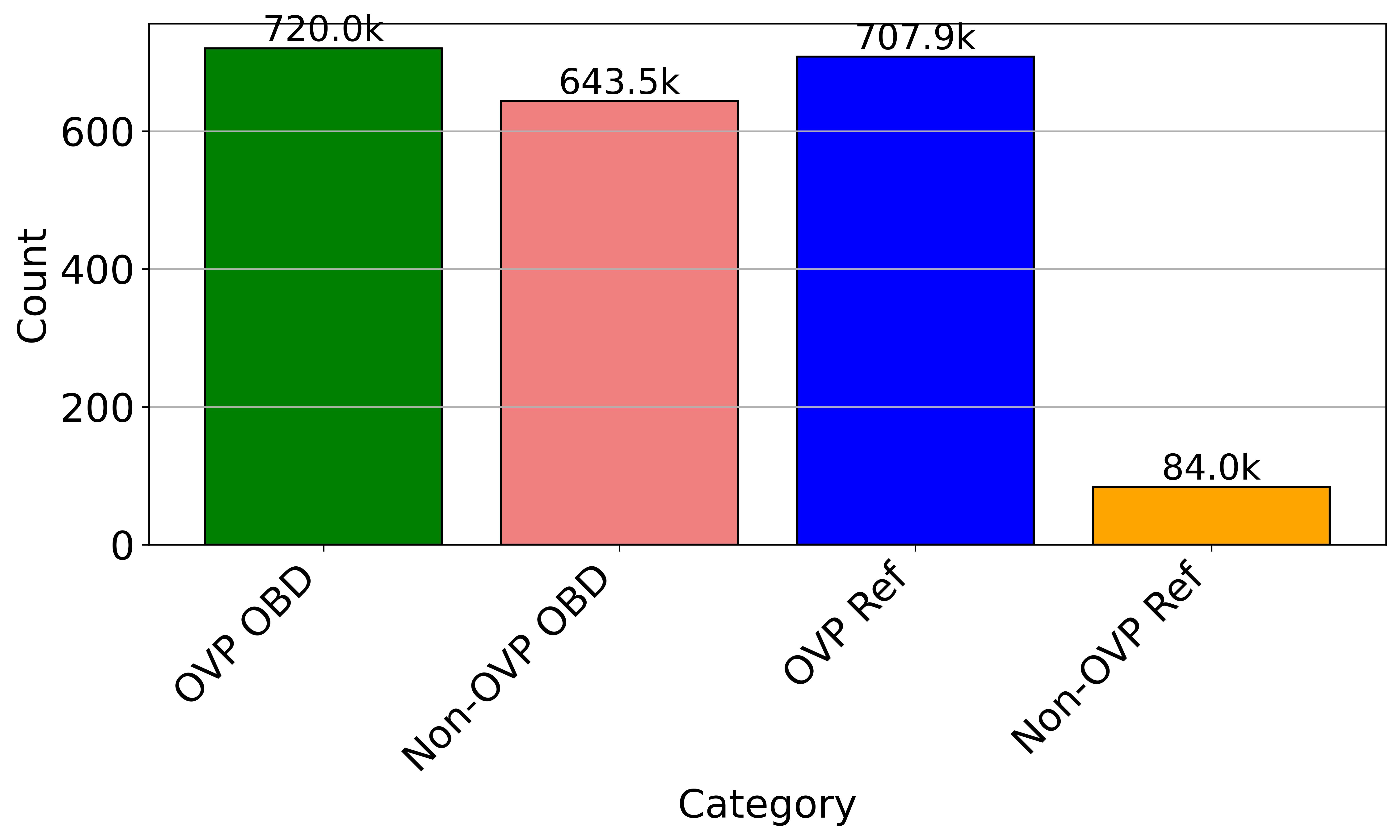} \\ (h) Houston --- Microsoft }
    \end{tabular}
    \caption{Distribution of the count of overlapping and non-overlapping buildings in study cities. Green represents overlapping OBD buildings, pink represents non-overlapping OBD buildings, blue represents overlapping reference buildings of OBD buildings, and gold represents non-overlapping reference buildings. Each row represents a city, with two columns displaying comparative distributions. }
    \label{fig:building_overlap_hist}
\end{figure}

\label{app1}

%%%%%%%%%%%%%%%%%%%%%%%%%%%%%%%%%%%%%%%%%%
%\isPreprints{}{% This command is only used for ``preprints''.
%\begin{adjustwidth}{-\extralength}{0cm}
%} % If the paper is ``preprints'', please uncomment this parenthesis.
%\printendnotes[custom] % Un-comment to print a list of endnotes

%\section{References}

% Please provide either the correct journal abbreviation (e.g. according to the “List of Title Word Abbreviations” http://www.issn.org/services/online-services/access-to-the-ltwa/) or the full name of the journal.
% Citations and References in Supplementary files are permitted provided that they also appear in the reference list here. 

%=====================================
% References, variant A: external bibliography
%=====================================

\bibliographystyle{apalike}   % or unsrt, ieeetr, apalike, etc.
\bibliography{reference}

% Chicago format (Used for journal: arts, genealogy, histories, humanities, jintelligence, laws, literature, religions, risks, socsci)

% If authors have biography, please use the format below
%\section*{Short Biography of Authors}
%\bio
%{\raisebox{-0.35cm}{\includegraphics[width=3.5cm,height=5.3cm,clip,keepaspectratio]{Definitions/author1.pdf}}}
%{\textbf{Firstname Lastname} Biography of first author}
%
%\bio
%{\raisebox{-0.35cm}{\includegraphics[width=3.5cm,height=5.3cm,clip,keepaspectratio]{Definitions/author2.jpg}}}
%{\textbf{Firstname Lastname} Biography of second author}

% For the MDPI journals use author-date citation, please follow the formatting guidelines on http://www.mdpi.com/authors/references
% To cite two works by the same author: \citeauthor{ref-journal-1a} (\citeyear{ref-journal-1a}, \citeyear{ref-journal-1b}). This produces: Whittaker (1967, 1975)
% To cite two works by the same author with specific pages: \citeauthor{ref-journal-3a} (\citeyear{ref-journal-3a}, p. 328; \citeyear{ref-journal-3b}, p.475). This produces: Wong (1999, p. 328; 2000, p. 475)

%%%%%%%%%%%%%%%%%%%%%%%%%%%%%%%%%%%%%%%%%%
%% for journal Sci
%\reviewreports{\\
%Reviewer 1 comments and authors’ response\\
%Reviewer 2 comments and authors’ response\\
%Reviewer 3 comments and authors’ response
%}
%%%%%%%%%%%%%%%%%%%%%%%%%%%%%%%%%%%%%%%%%%

\end{document}